\documentclass[12pt, epsfig, twocolumn]{aastex6}
\usepackage{graphicx}
\usepackage{color}
\usepackage{epstopdf}

\shorttitle{Local Group distances and publication bias. VII.}
\shortauthors{Richard de Grijs and Giuseppe Bono}

\begin{document}

\title{Clustering of Local Group distances: publication bias or
  correlated measurements? VII. A distance framework out to 100 Mpc}

\author{
Richard de Grijs\altaffilmark{1,2,3} and
Giuseppe Bono\altaffilmark{4,5}
}

\altaffiltext{1} {Department of Physics \& Astronomy, Macquarie
  University, Balaclava Road, Sydney, NSW 2109, Australia}
\altaffiltext{2} {Research Centre for Astronomy, Astrophysics \&
  Astrophotonics, Macquarie University, Balaclava Road, Sydney, NSW
  2109, Australia}
\altaffiltext{3} {International Space Science Institute--Beijing, 1
  Nanertiao, Zhongguancun, Hai Dian District, Beijing 100190, China}
\altaffiltext{4} {Dipartimento di Fisica, Universit\`a di Roma Tor
  Vergata, via Della Ricerca Scientifica 1, 00133, Roma, Italy}
\altaffiltext{5} {INAF, Rome Astronomical Observatory, via Frascati
  33, 00078 Monte Porzio Catone, Italy}

\begin{abstract}
We consider the body of published distance moduli to the Fornax and
Coma galaxy clusters, with specific emphasis on the period since
1990. We have carefully homogenized our final catalogs of distance
moduli onto the distance scale established in the previous papers in
this series. We assessed systematic differences associated with the
use of specific tracers, and consequently discarded results based on
application of the Tully--Fisher relation and of globular cluster and
planetary nebula luminosity functions. We recommend `best' weighted
relative distance moduli for the Fornax and Coma clusters with respect
to the Virgo cluster benchmark of $\Delta (m-M)_0^{\rm Fornax - Virgo}
= 0.18 \pm 0.28 $ mag and $\Delta (m-M)_0^{\rm Coma - Virgo} = 3.75
\pm 0.23$ mag. The set of weighted mean distance moduli (distances) we
derived as most representative of the clusters' distances is,
\begin{eqnarray}
(m-M)_0^{\rm Fornax} &=& 31.41 \pm 0.15 \mbox{ mag } (D = 19.1^{+1.4}_{-1.2} \mbox{ Mpc) and} \nonumber \\
                     &=& 31.21 \pm 0.28 \mbox{ mag } (D = 17.5^{+2.4}_{-2.2} \mbox{ Mpc)}; \nonumber \\
(m-M)_0^{\rm Coma}   &=& 34.99 \pm 0.38 \mbox{ mag } (D = 99.5^{+19.0}_{-15.9} \mbox{ Mpc) and} \nonumber \\
                     &=& 34.78 \pm 0.27 \mbox{ mag } (D = 90.4^{+11.9}_{-10.6} \mbox{ Mpc)}, \nonumber
\end{eqnarray}
where the first distance modulus for each cluster is the result of our
analysis of the direct, absolute distance moduli, while the second
modulus is based on distance moduli relative to the Virgo
cluster. While the absolute and relative distance moduli for both
clusters are mutually consistent within the uncertainties, the
relative distance moduli yield shorter absolute distances by
$\sim$1$\sigma$. It is unclear what may have caused this small
difference for both clusters; lingering uncertainties in the
underlying absolute distance scale appear to have given rise to a
systematic uncertainty on the order of 0.20 mag.
\end{abstract}


\section{A robust framework of extragalactic distances}

Over the course of the past decade, we have established a robust
distance framework to galaxies in the Local Group and beyond, based on
a set of mutually and internally consistent distance moduli that were
validated on a robust statistical basis. In \citet[][henceforth Paper
  I]{2014AJ....147..122D}, we explored the presence of `publication
bias' in the body of published distance determinations for the Large
Magellanic Cloud (LMC), as suggested by
\citet{2008AJ....135..112S}. While we did not find any evidence of
authors have jumped on this proverbial bandwagon, we put
\citet{2001ApJ...553...47F}'s canonical LMC distance modulus of
$(m-M)_0^{\rm LMC} = 18.50 \pm 0.10$ mag on a well-established
statistical footing, recommending $(m-M)_0^{\rm LMC} = 18.49 \pm 0.09$
mag \citep[see also][]{2015ApJ...815...87C}.

This was followed by a series of papers aimed at both exploring the
reality and/or the presence of publication bias among published
distance measurements and establishing a robust local distance
framework. In order of increasing distance, we applied the same
analysis as developed in Paper I to the Galactic Center \citep[][Paper
  IV]{2016ApJS..227....5D} and the Galactic rotation constants
\citep[][Paper V]{2017ApJS..232...22D}, the Small Magellanic Cloud
\citep[][Paper III; see also
  \citealt{2015ApJ...815...87C}]{2015AJ....149..179D}, the M31 group
\citep[][Paper II]{2014AJ....148...17D}, and the Virgo cluster
\citep[][Paper VI]{2020ApJS..246....3D}. In addition, we strongly
recommend the independent, geometric distannce measurement to the
maser host galaxy NGC 4258 published by \citet{1999Natur.400..539H} as
additional, intermediate-distance stepping stone. Our full, internally
consistent distance framework thus far established is summarized in
Table 3 of Paper VI.

In this paper, we expand our previous analyses by focusing on two
additional, rich benchmark galaxy clusters. These include the Fornax
cluster as southern-hemisphere benchmark counterpart to the Virgo
cluster in the northern hemisphere, as well as the Coma cluster. This
takes our internally consistent `local' distance framework out to
distances of order 100 Mpc. At those distances, a significant fraction
(although still a large minority) of articles citing distance
estimates refer to redshifts rather than linear scales. This will
therefore conclude our efforts to establish a benchmark set of
statistically validated distance estimates in the local Universe.

We have organized this paper as follows. In Section \ref{data.sec} we
briefly summarize our data-mining approach and describe the resulting
catalogs containing Fornax and Coma distances. We examine trends in
distance determinations for both clusters in Section
\ref{stats.sec}. Then, in Section \ref{trends.sec}, we analyze the
systematic differences, if any, among tracer populations for both
clusters. This eventually results in a set of recommended benchmark
distances, which we summarize in Section \ref{summary.sec}. These
should be combined with the distance moduli summarized in Paper VI
(see that paper's Table 3) to gain a full overview of our
extragalactic distance framework.

\section{Our database}
\label{data.sec}

Similarly to the previous papers in this series, we mined the
NASA/Astrophysics Data System (ADS) for newly derived or
recalibrated/updated distance measures to the Fornax and Coma
clusters. We used as search terms `Fornax Cluster' and `Coma Cluster.'
We adopted the same criteria as in our previous papers for inclusion
of any new values in our final database (for a detailed description,
see, e.g., Paper I). We included both measurements to the galaxy
clusters as a whole, as well as to individual galaxies in the cluster
cores (indicated separately in our final catalogs). For the Fornax
cluster, we hence included distance measures to NGC 1316, NGC 1326A,
NGC 1365, and NGC 1399 (as well as NGC 1404). The relevant galaxies in
the Coma cluster for which distance moduli are included in our final
database are NGC 4874, NGC 4881, NGC 4889, NGC 4921, NGC 4923, and IC
4051.

For the `modern' period, from 1990 onward, we carefully perused all
articles resulting from our NASA/ADS queries; prior to 1990, we
followed the reference trail. As of 2019 December 19, when we
completed our data mining, the numbers of hits in the NASA/ADS for the
Fornax and Coma clusters were 1849 and 5357, respectively. The
resulting numbers of absolute/relative distance measures included in
our final catalogs are 140/62 and 95/56 for the Fornax and Coma
clusters, respectively. For the Fornax cluster, we retrieved distance
moduli relative to that of the Virgo cluster; for the Coma cluster, we
retrieved relative distance measures with respect to the Virgo (53),
Leo I (2 direct, 10 indirect; see Section \ref{relative.sec}), and
Fornax (1) clusters.

Our final database, sorted by year and by tracer for both galaxy
clusters separately, is available online through
http://astro-expat.info/Data/pubbias.html,\footnote{A permanent link
  to this page can be found at
  http://web.archive.org/web/20200331174040/http://astro-expat.info/Data/pubbias.html;
  members of the community are encouraged to send us updates or
  missing information.} For graphical depictions of the clusters'
distance moduli as a function of year of publication, see Figures
\ref{fig1} and \ref{fig2}. We will discuss the trends and any evidence
of publication bias versus correlated measures in Section
\ref{trends.sec}.

\section{Trends in Fornax and Coma Cluster distance determinations}
\label{trends.sec}

\begin{figure*}[ht!]
\plotone{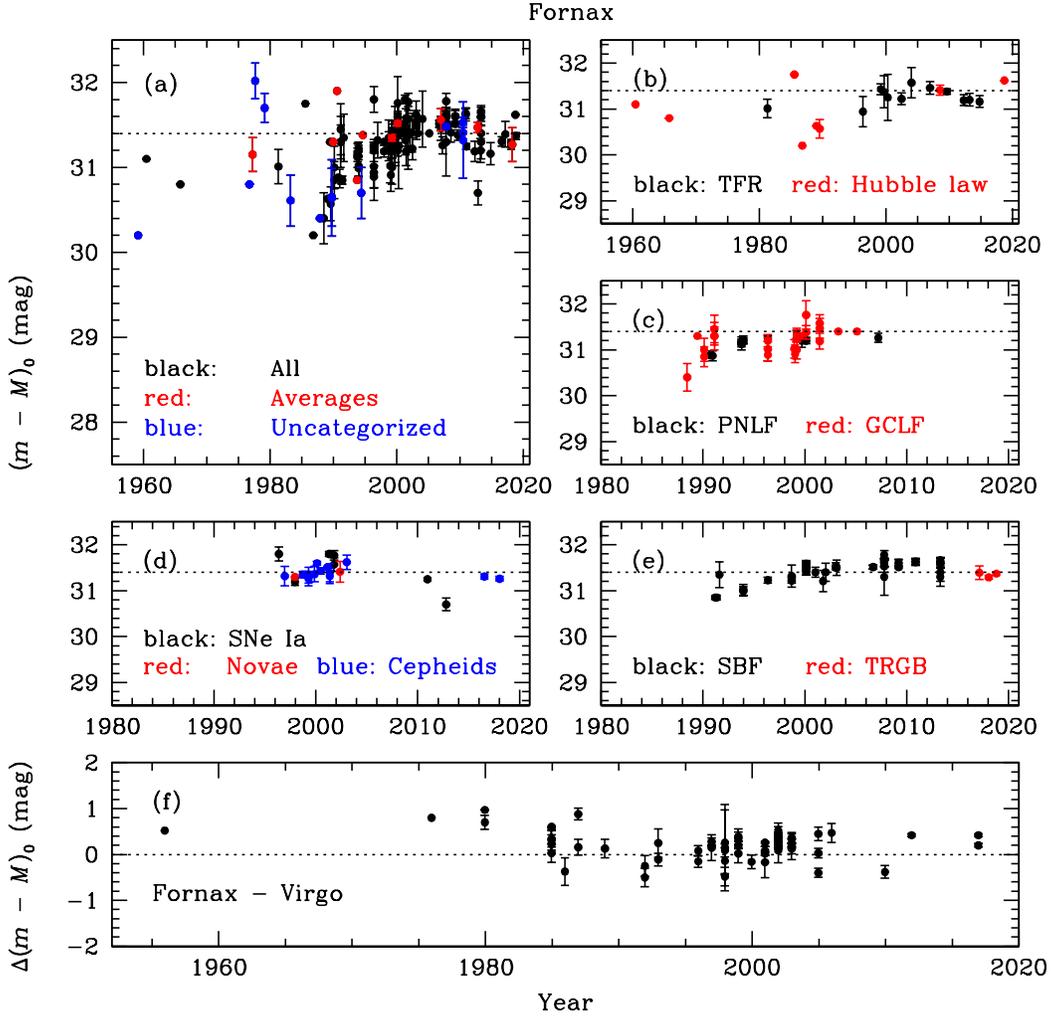}
\vspace*{-3cm}
\caption{Published distance moduli (original values and original error
  bars, where available) to the Fornax cluster and its central
  galaxies, NGC 1316, NGC 1326A, NGC 1365, and NGC 1399 (as well as
  NGC 1404). The horizontal dotted lines in panels (a)--(e) represent
  $(m-M)_0 = 31.4$ mag and are meant to guide the eye. GCLF: Globular
  cluster luminosity function; PNLF: Planetary nebula luminosity
  function; TFR: Tully--Fisher relation; TRGB: Tip of the red-giant
  branch; SBF: Surface brightness fluctuations; SNe Ia: Type Ia
  supernovae. `Averages' in panel (a) include weighted and unweighted
  means of different methods of distance determination, as well as
  mean values of the distance moduli to samples of central Fornax
  cluster galaxies, as published by the original authors (see Section
  10 in our externally linked data table); `Uncategorized' distance
  moduli include any measurements that are not already included in the
  other panels, mostly because of the scarcity of data points for a
  particular measurement approach (see Section 11 of the same
  table). Panel (f) shows the set of published relative Fornax--Virgo
  Cluster distance moduli (any tracer), where positive values reflect
  a greater distance to the Fornax cluster compared with Virgo.}
\label{fig1}
\end{figure*}

\begin{figure*}[ht!]
\plotone{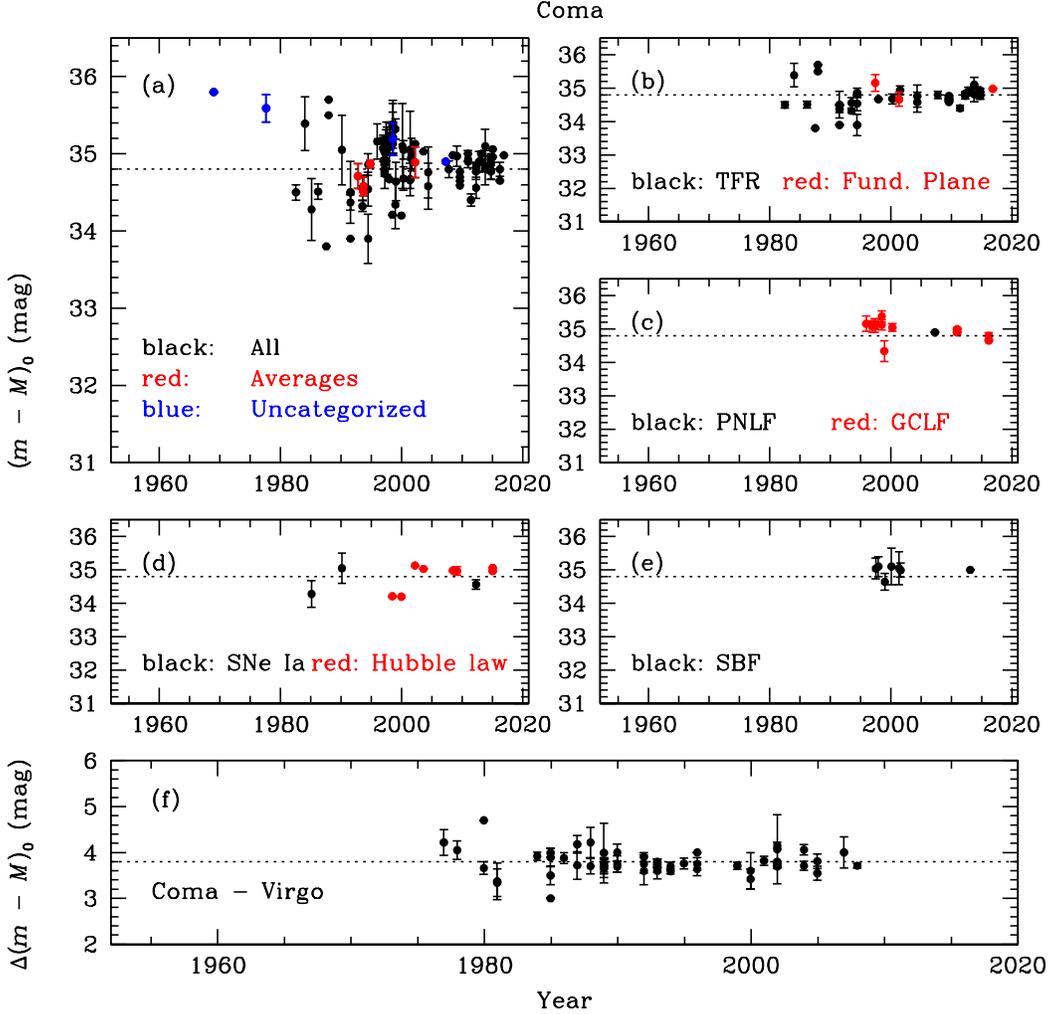}
\vspace*{-3cm}
\caption{As Figure \ref{fig1}, but for the Coma cluster and its
  central galaxies, NGC 4874, NGC 4881, NGC 4889, NGC 4921, NGC 4923,
  and IC 4051. Fund. Plane: Fundamental Plane. Panel (f) shows the
  relative Coma--Virgo cluster distance moduli (as in Figure
  \ref{fig1}f).}
\label{fig2}
\end{figure*}

Figures \ref{fig1} and \ref{fig2} show both the overall distribution
of distance moduli and the distance measures for selected individual
tracers for the Fornax and Coma clusters, respectively. The bottom
panels in both figures show the sets of published relative distance
moduli of our target clusters with respect to the Virgo cluster.

Since we have focused our detailed data mining on the `modern' period
from 1990 onward, we will concentrate on those measures in our
analysis of the distances implied by the individual tracers. However,
careful assessment of both figures shows that hidden trends in the
distance moduli may be present for Fornax distance measures based on
application of the Tully--Fisher relation (TFR), surface brightness
fluctuations (SBF), and the planetary nebula luminosity function
(PNLF). Hence, for the Fornax distances based on the TFR and SBF, we
analyzed the period from 2000 onward, while for PNLF-based measures we
used the timeframe since 1995. For the Coma cluster, we restricted our
analysis of its TFR-based distances to the period from 2000 given the
large scatter of the individual data points prior to 2000, which may
imply the presence of unaccounted-for biases and measurement
errors. Note that all data points included in Figures \ref{fig1} and
\ref{fig2} reflect the assumptions made by their original authors. In
Section \ref{stats.sec} we will first homogenize the data points used
for further analysis before drawing our conclusions.

Although we retain the Fornax distance modulus suggested by
\citet{2001ApJ...548L.139D} in our online catalog
(http://astro-expat.info/Data/pubbias.html), we did not include this
measurement in our analysis. It is based on the mean distance modulus
resulting from three Cepheid-based measurements available at the time
of publication, to NGC 1365, NGC1326A, NGC 1425. These authors suggest
that neither NGC 1326A nor NGC 1425 may be representative of the
cluster core (but note that our final catalog retains the Cepheid
distance to NGC 1326A, since it does not appear out of place with
respect to the overall body of measurements), thus only leaving the
Cepheid distance to NGC 1365 \citep{1999ApJ...515...29M}. The latter
is already included in our Fornax catalog, thus rendering the
\citet{2001ApJ...548L.139D} result superfluous.

Similarly, we retain two distances based on globular cluster
luminosity functions (GCLFs) in our final catalog of Fornax cluster
distance measures, which we however do not use in our analysis. This
relates to the distance moduli of \citet{1997eds..proc..254W} and
\citet{2003AJ....125.1908D}. The former reference does not provide
sufficient information for us to include its distance determination in
our analysis; the latter distance determination is, in essence, based
on an assessment of the measured peak of the GCLF as being
`consistent' with the SBF-based distance measurements of
\citet{2001ApJ...546..681T} and \citet{2002ApJ...564..216L}. As such,
it is not a firm determination.

\section{Statistically validated distances}
\label{stats.sec}

\subsection{Systematic differences affecting individual tracers}

In the previous section, we identified a number of individual distance
tracers for both galaxy clusters for which we had collected a
sufficient number of measurements to analyze their distance
distributions separately. We show their distance measures separately
in panels (b)--(e) in both Figures \ref{fig1} and \ref{fig2}. Table
\ref{tab1} includes the weighted means for each tracer and for the
relevant period of interest, as well as the relevant 1$\sigma$
uncertainties. 

\begin{figure*}[ht!]
\plotone{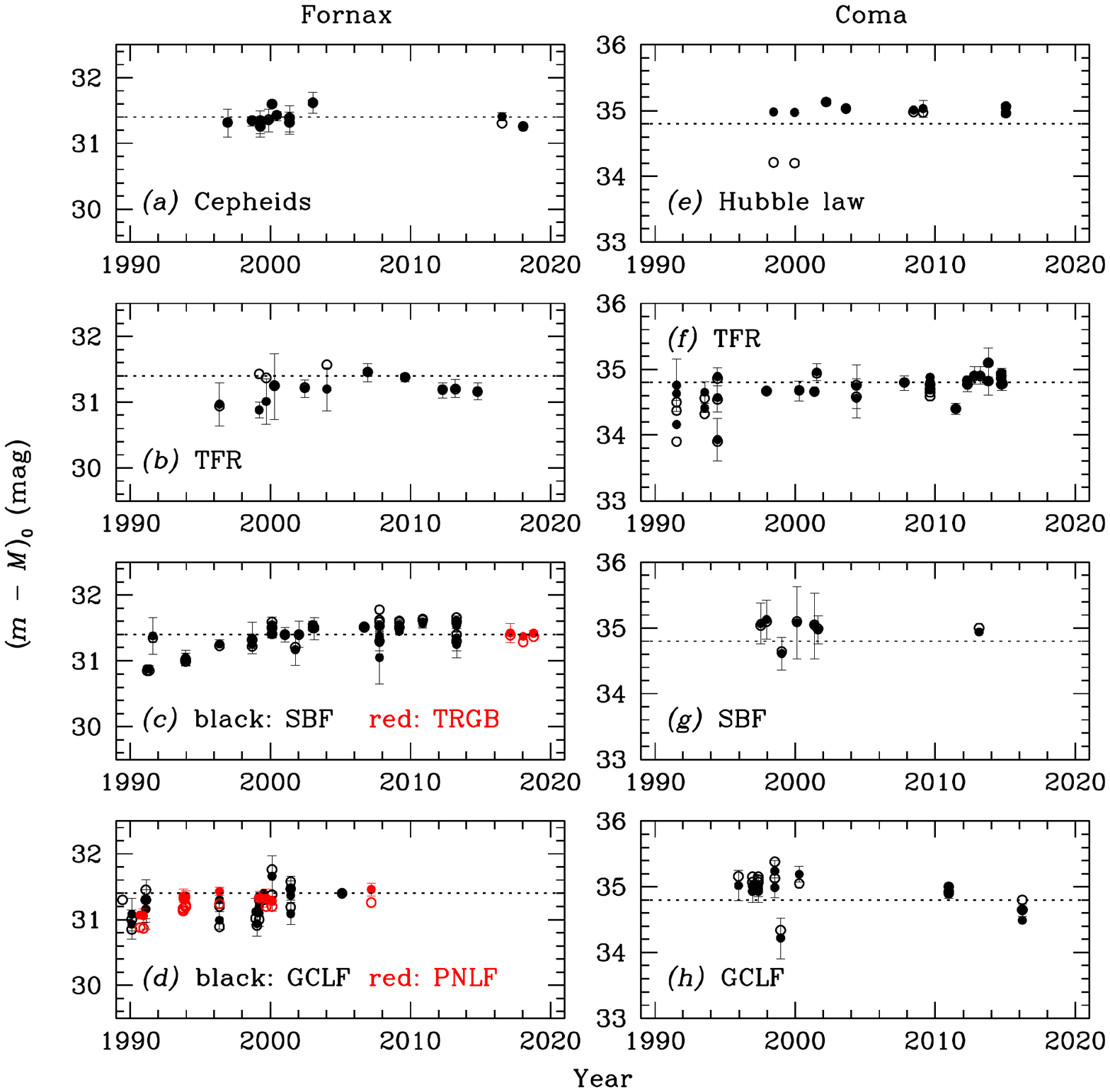}
\vspace*{-3cm}
\caption{Original and homogenized distance moduli published since 1990
  to (left; a--d) the Fornax and (right; e--h) the Coma clusters for
  specific tracers. The horizontal dotted lines, meant to guide the
  eye, are drawn at $(m-M)_0 = 31.4$ and 34.8 mag for the Fornax and
  Coma clusters, respectively. Open circles represent original,
  published values; solid bullets have been corrected to a common
  distance framework (see text). Error bars are included where they
  were provided by the original authors.}
\label{fig3}
\end{figure*}

However, before compiling Table \ref{tab1}, we ensured that all
distance moduli contributing to the weighted means were carefully
homogenized onto our overall distance framework as determined in this
series of papers. Tables \ref{tab2} and \ref{tab3} include the
numerical basis of this homogenization, showing both the zero-point
calibrations used by the original authors and our adjustments of their
distance moduli to match our distance framework thus far established
(see footnote {\it a} to Table \ref{tab2} for a quick overview). The
homogenized distance moduli (as well as the original values) are
displayed as a function of publication date and for each of the main
tracers separately in Figure \ref{fig3}.

\begin{table*}
\vspace{2cm}
\caption{Mean, post-1990 published distance measures to the centers of
  the Fornax and Coma Clusters as a function of tracer population.}
\begin{center}
\label{tab1}
\begin{tabular}{@{}lcrcclcrc@{}}
\hline \hline
\multicolumn{4}{c}{Fornax} & & \multicolumn{4}{c}{Coma} \\
\cline{1-4}\cline{6-9}
\multicolumn{1}{l}{Tracer} & \multicolumn{1}{c}{Period} &
\multicolumn{1}{r}{$N$} & \multicolumn{1}{c}{$(m-M)_0$} & &
\multicolumn{1}{l}{Tracer} & \multicolumn{1}{c}{Period} &
\multicolumn{1}{r}{$N$} & \multicolumn{1}{c}{$(m-M)_0$} \\
& & & (mag) & & & & & (mag) \\
\hline
Cepheids$^*$ & 1990--2019 & 13 & 31.38 $\pm$ 0.14 & & Hubble law$^a$ & 1990--2019 &  8 & 35.02 $\pm$ 0.06 \\
TFR          & 2000--2019 &  8 & 31.25 $\pm$ 0.24 & & TFR$^*$        & 2000--2019 & 33 & 34.72 $\pm$ 0.18 \\
SBF          & 2000--2019 & 28 & 31.44 $\pm$ 0.19 & & SBF$^*$        & 1990--2019 &  7 & 34.98 $\pm$ 0.37 \\
TRGB         & 1990--2019 &  3 & 31.41 $\pm$ 0.09 & & GCLF$^*$       & 1990--2019 & 15 & 34.90 $\pm$ 0.17 \\
GCLF$^*$     & 1990--2019 & 19 & 31.22 $\pm$ 0.21 \\
PNLF         & 1995--2019 &  5 & 31.36 $\pm$ 0.09 \\ 
\hline \hline
\end{tabular}
\end{center}
\flushleft 
$^a$ Adopting $H_0 = 70$ km s$^{-1}$ Mpc$^{-1}$; a reduction
(increase) to $H_0 = 67.3$ (72) km s$^{-1}$ Mpc$^{-1}$ (spanning the
range implied by the current `Hubble tension') would result in an
increase (decrease) in the Coma distance modulus by 0.09 (0.06)
mag. The uncertainty represents the Gaussian width ($\sigma$) of the
distribution, given that none of the published Coma distances based on
application of the Hubble law included estimates of the associated
uncertainties.\\
$^*$ Since a number of published distance measures did not include
associated uncertainties, the central values are based on the full set
of published measurements, while the uncertainties only include those
values that were published with their associated uncertainties. This
affects the following numbers of measurements: (1) Fornax -- Cepheids:
1; GCLF: 3; (2) Coma -- GCLF: 3; SBF: 1; TFR: 8.
\end{table*}

\begin{table*}
\vspace{2cm}
\caption{Corrections to published Fornax cluster distance moduli.}
\begin{center}
\label{tab2}
\tabcolsep 0.5mm
{\tiny
\begin{tabular}{@{}cclcccll@{}}
\hline \hline
\multicolumn{1}{c}{Date} & \multicolumn{1}{c}{$(m-M)_0^{\rm orig}$} &
\multicolumn{1}{c}{Orig. calibration} & \multicolumn{1}{c}{Correction$^a$} &
\multicolumn{1}{c}{$(m-M)_0^{\rm scaled}$} &
\multicolumn{1}{c}{Target} & \multicolumn{1}{c}{Notes} &
\multicolumn{1}{c}{Reference} \\
\multicolumn{1}{c}{(yyyy/mm)} & \multicolumn{1}{c}{(mag)} & &
\multicolumn{1}{c}{(mag)} & \multicolumn{1}{c}{(mag)} \\
\hline 
\multicolumn{8}{c}{Cepheids} \\
\hline
1996/12 & $31.32 \pm 0.21$ & $(m-M)_0^{\rm LMC}   = 18.50$  mag & $-0.01$ & 31.31 & NGC 1365 & & \citet{1996AAS...18910804M} \\ 
1998/09 & $31.35 \pm 0.07$ & $(m-M)_0^{\rm LMC}   = 18.50$  mag & $-0.01$ & 31.34 & NGC 1365 & & \citet{1998Natur.395...47M} \\ 
1999/04 & $31.35 \pm 0.07$ & $(m-M)_0^{\rm LMC}   = 18.50$  mag & $-0.01$ & 31.34 & NGC 1365 & & \citet{1999ApJ...515...29M} \\ 
1999/04 & $31.31 \pm 0.20$ & $(m-M)_0^{\rm LMC}   = 18.50$  mag & $-0.01$ & 31.30 & NGC 1365 & & \citet{1999ApJ...515....1S} \\ 
1999/04 & $31.26 \pm 0.10$ & $(m-M)_0^{\rm LMC}   = 18.50$  mag & $-0.01$ & 31.25 & NGC 1365 & & \citet{1999ApJ...515....1S} \\ 
1999/11 & $31.36 \pm 0.17$ & $(m-M)_0^{\rm LMC}   = 18.50$  mag & $-0.01$ & 31.35 & NGC 1326A& & \citet{1999ApJ...525...80P} \\ 
2000/02 & $31.60 \pm 0.04$ & $(m-M)_0^{\rm LMC}   = 18.50$  mag & $-0.01$ & 31.59 & Fornax   & & \citet{2000ApJ...529..745F} \\ 
2000/06 & $31.43 \pm 0.07$ & $(m-M)_0^{\rm LMC}   = 18.50$  mag & $-0.01$ & 31.42 & NGC 1365 & & \citet{2000ApJ...529..745F} \\ 
2001/05 & $31.32 \pm 0.17$ & $(m-M)_0^{\rm LMC}   = 18.50$  mag & $-0.01$ & 31.31 & Fornax   & & \citet{2001ApJ...553...47F} \\ 
2001/05 & $31.39 \pm 0.20$ & $(m-M)_0^{\rm LMC}   = 18.50$  mag & $-0.01$ & 31.38 & Fornax   & Corr. for metallicity & \citet{2001ApJ...553...47F} \\ 
2003/01 & $31.62 \pm 0.16$ & $M_I^{\rm TRGB}      = -4.05$  mag & $ 0.00$ & 31.62 & Fornax   & & \citet{2003AnA...398...63J} \\ 
2016/07 & $31.307\pm 0.057$& $(m-M)_0^{\rm N4258} = 29.387$ mag & $-0.10$ & 31.21 & NGC 1365 & & \citet{2016ApJ...826...56R} \\ 
2018/01 & $31.26 \pm 0.05$ & $(m-M)_0^{\rm LMC}   = 18.49$  mag & $ 0.00$ & 31.26 & NGC 1365 & & \citet{2018ApJ...852...60J} \\ 
\hline 
\multicolumn{8}{c}{SBF} \\
\hline
1991/03 & $30.85 \pm 0.04$ & $(m-M)_0^{\rm M31}   = 24.43$ mag & $+0.03$ & 30.88 & Fornax   & & \citet{1991BAAS...23..956T} \\ 
1991/05 & $30.85 \pm 0.05$ & $(m-M)_0^{\rm M31}   = 24.43$ mag & $+0.03$ & 30.88 & Fornax   & & \citet{1991ApJ...373L...1T} \\ 
1991/08 & $31.35 \pm 0.28$ & $(m-M)_0^{\rm M31}   = 24.43$ mag & $+0.03$ & 31.38 & Fornax   & & \citet{1991ApJ...376..404B} \\ 
1993/12 & $31.02 \pm 0.12$ & $(m-M)_0^{\rm M31}   = 24.43$ mag & $+0.03$ & 31.05 & NGC 1316 & & \citet{1993ApJ...419..479C} \\ 
1993/12 & $30.99 \pm 0.10$ & $(m-M)_0^{\rm M31}   = 24.43$ mag & $+0.03$ & 31.02 & NGC 1399 & & \citet{1993ApJ...419..479C} \\ 
1996/05 & $31.23 \pm 0.06$ & $(m-M)_0^{\rm M31}   = 24.43$ mag & $+0.03$ & 31.26 & Fornax   & & \citet{1997eds..proc..297T} \\ 
1998/09 & $31.22 \pm 0.06$ & $(m-M)_0^{\rm Virgo} = 31.00$ mag & $+0.03$ & 31.25 & Fornax   & $I$ & \citet{1998ApJ...505..111J} \\ 
1998/09 & $31.32 \pm 0.24$ & $(m-M)_0^{\rm Virgo} = 31.00$ mag & $+0.03$ & 31.35 & Fornax   & $K'$ & \citet{1998ApJ...505..111J} \\ 
2000/02 & $31.41 \pm 0.06$ & $^b$                              & $+0.01$ & 31.42 & Fornax   & $I$ & \citet{2000ApJ...530..625T} \\ 
2000/02 & $31.50 \pm 0.16$ & $^b$                              & $+0.01$ & 31.51 & NGC 1399 & $I$ & \citet{2000ApJ...530..625T} \\ 
2000/02 & $31.59 \pm 0.04$ & $+0.05$ mag w.r.t. \citet{2001ApJ...546..681T} & $-0.04$ & 31.55 & Fornax & $I$ & \citet{2000ApJ...529..745F} \\ 
2000/02 & $31.51 \pm 0.08$ & $+0.05$ mag w.r.t. \citet{2001ApJ...546..681T} & $-0.04$ & 31.47 & Fornax & $K'$ & \citet{2000ApJ...529..745F} \\ 
2001/01 & $31.40 \pm 0.11$ & $(m-M)_0^{\rm Virgo} = 31.03$ mag & $ 0.00$ & 31.40 & Fornax   & $I$ & \citet{2001MNRAS.320..193B} \\ 
2001/10 & $31.21 \pm 0.23$ & based on \citet{2000ApJ...529..745F} & $-0.04$ & 31.17 & NGC 1316 & & \citet{2001ApJ...559..584A} \\ 
2002/01 & $31.4  \pm 0.2 $ & \citet{2000ApJ...530..625T} calibration   & $+0.01$ & 31.41 & NGC 1399 & & \citet{2002ApJ...564..216L} \\ 
2003/01 & $31.54 \pm 0.07$ & $M_I^{\rm TRGB}      = -4.05$ mag & $ 0.00$ & 31.54 & Fornax   & Dwarf galaxies & \citet{2003AnA...398...63J} \\ 
2003/01 & $31.50 \pm 0.04$ & $M_I^{\rm TRGB}      = -4.05$ mag & $ 0.00$ & 31.50 & Fornax   & Early-type galaxies & \citet{2003AnA...398...63J} \\ 
2003/02 & $31.50 \pm 0.17$ & $(m-M)_0^{\rm LMC}   = 18.50$ mag & $-0.01$ & 31.49 & NGC 1316 & & \citet{2003ApJ...583..712J} \\ 
2006/09 & $31.51 \pm 0.04$ & $M_I^{\rm TRGB}      = -4.05$ mag & $ 0.00$ & 31.51 & Fornax   & & \citet{2006AJ....132.1384D} \\ 
2006/09 & $31.52 \pm 0.04$ & $M_I^{\rm TRGB}      = -4.05$ mag & $ 0.00$ & 31.52 & NGC 1399 & & \citet{2006AJ....132.1384D} \\ 
2007/10 & $31.78 \pm 0.09$ & $-0.21$ mag w.r.t. \citet{2000ApJ...529..745F} & $-0.25$ & 31.53 & NGC 1316 & & \citet{2007ApJ...668..130C} \\ 
2007/10 & $31.62 \pm 0.09$ & $-0.21$ mag w.r.t. \citet{2000ApJ...529..745F} & $-0.25$ & 31.37 & NGC 1316 & & \citet{2007ApJ...668..130C} \\ 
2007/10 & $31.62 \pm 0.09$ & $-0.21$ mag w.r.t. \citet{2000ApJ...529..745F} & $-0.25$ & 31.37 & NGC 1316 & & \citet{2007ApJ...668..130C} \\ 
2007/10 & $31.53 \pm 0.13$ & $-0.21$ mag w.r.t. \citet{2000ApJ...529..745F} & $-0.25$ & 31.28 & NGC 1316 & & \citet{2007ApJ...668..130C} \\ 
2007/10 & $31.59 \pm 0.08$ & $-0.21$ mag w.r.t. \citet{2000ApJ...529..745F} & $-0.25$ & 31.34 & NGC 1316 & & \citet{2007ApJ...668..130C} \\ 
2007/10 & $31.3  \pm 0.4 $ & $-0.21$ mag w.r.t. \citet{2000ApJ...529..745F} & $-0.25$ & 31.05 & Fornax   & & \citet{2007ApJ...668..130C} \\ 
2009/03 & $31.51 \pm 0.03$ & $(m-M)_0^{\rm Virgo} = 31.09$ mag & $-0.06$ & 31.45 & Fornax   & $z$ & \citet{2009ApJ...694..556B} \\ 
2009/03 & $31.606\pm 0.065$& $(m-M)_0^{\rm Virgo} = 31.09$ mag & $-0.06$ & 31.55 & NGC 1316 & $z$ & \citet{2009ApJ...694..556B} \\ 
2009/03 & $31.596\pm 0.091$& $(m-M)_0^{\rm Virgo} = 31.09$ mag & $-0.06$ & 31.54 & NGC 1399 & $z$ & \citet{2009ApJ...694..556B} \\  
2010/11 & $31.620\pm 0.071$& $+0.06$ mag w.r.t. \citet{2001ApJ...546..681T} & $-0.05$ & 31.57 & NGC 1399 & F814W & \citet{2010ApJ...724..657B} \\ 
2010/11 & $31.638\pm 0.066$& $+0.06$ mag w.r.t. \citet{2001ApJ...546..681T} & $-0.05$ & 31.59 & NGC 1316 & F814W & \citet{2010ApJ...724..657B} \\ 
2013/04 & $31.59 \pm 0.05$ & $+0.06$ mag w.r.t. \citet{2001ApJ...546..681T} & $-0.05$ & 31.54 & NGC 1316 & & \citet{2013AnA...552A.106C} \\ 
2013/04 & $31.60 \pm 0.11$ & $+0.06$ mag w.r.t. \citet{2001ApJ...546..681T} & $-0.05$ & 31.55 & NGC 1316 & & \citet{2013AnA...552A.106C} \\ 
2013/04 & $31.66 \pm 0.07$ & $+0.06$ mag w.r.t. \citet{2001ApJ...546..681T} & $-0.05$ & 31.61 & NGC 1316 & $z$ & \citet{2013AnA...552A.106C} \\ 
2013/04 & $31.3  \pm 0.2 $ & $+0.06$ mag w.r.t. \citet{2001ApJ...546..681T} & $-0.05$ & 31.25 & NGC 1316 & $z$ & \citet{2013AnA...552A.106C} \\ 
2013/04 & $31.4  \pm 0.2 $ & $+0.06$ mag w.r.t. \citet{2001ApJ...546..681T} & $-0.05$ & 31.35 & NGC 1316 & $z$ & \citet{2013AnA...552A.106C} \\ 
\hline
\hline 
\end{tabular}
}
\end{center}
\end{table*}

\addtocounter{table}{-1}
\begin{table*}
\vspace{2cm}
\caption{(Continued)}
\begin{center}
\tabcolsep 0.5mm
{\tiny
\begin{tabular}{@{}cclcccll@{}}
\hline \hline
\multicolumn{1}{c}{Date} & \multicolumn{1}{c}{$(m-M)_0^{\rm orig}$} &
\multicolumn{1}{c}{Orig. calibration} & \multicolumn{1}{c}{Correction$^a$} &
\multicolumn{1}{c}{$(m-M)_0^{\rm scaled}$} &
\multicolumn{1}{c}{Target} & \multicolumn{1}{c}{Notes} &
\multicolumn{1}{c}{Reference} \\
\multicolumn{1}{c}{(yyyy/mm)} & \multicolumn{1}{c}{(mag)} & &
\multicolumn{1}{c}{(mag)} & \multicolumn{1}{c}{(mag)} \\
\hline 
\multicolumn{8}{c}{GCLF} \\
\hline
1990/02 & $31.0  \pm 0.25$ & $R_0 = 8.0$ kpc                       & $+0.08$ & 31.08 & Fornax   & & \citet{1990ApJ...350L...5G} \\ 
1990/02 & $30.85 \pm 0.22$ & $R_0 = 8.0$ kpc                       & $+0.08$ & 30.93 & NGC 1399 & & \citet{1990ApJ...350L...5G} \\ 
1991/02 & $31.3  \pm 0.2 $ & $M_{V,{\rm TO}}^{\rm MW} = -7.36$ mag & $-0.14$ & 31.16 & NGC 1399 & & \citet{1991AJ....101..469B} \\ 
1991/02 & $31.45 \pm 0.30$ & $M_{V,{\rm TO}}^{\rm MW} = -7.36$ mag & $-0.14$ & 31.31 & Fornax   & & \citet{1991AJ....101..469B} \\ 
1991/02 & $31.3  \pm 0.3 $ & $M_{V,{\rm TO}}^{\rm MW} = -7.36$ mag & $-0.14$ & 31.16 & Fornax   & & \citet{1991AJ....101..469B} \\ 
1996/05 & $31.20 \pm 0.13$ & $M_{V,{\rm TO}}^{\rm MW} = -7.40$ mag & $+0.10$ & 31.30 & Fornax   & $V$ & \citet{1996AnA...309L..39K} \\ 
1996/05 & $30.89 \pm 0.13$ & $M_{V,{\rm TO}}^{\rm MW} = -7.40$ mag & $+0.10$ & 30.99 & Fornax   & $I$ & \citet{1996AnA...309L..39K} \\ 
1998/12 & $31.02 \pm 0.2 $ & $M_{V,{\rm TO}}^{\rm MW} = -7.40$ mag & $+0.10$ & 31.12 & NGC 1399 & & \citet{1998AJ....116.2854O} \\ 
1999/01 & $30.91 \pm 0.19$ & $(m-M)_0^{\rm M31}       = 24.43$ mag & $+0.03$ & 30.94 & NGC 1399 & & \citet{1999AJ....117..167G} \\ 
1999/03 & $31.30 \pm 0.13$ & $M_{V,{\rm TO}}^{\rm MW} = -7.61$ mag & $-0.11$ & 31.19 & Fornax   & & \citet{2000fepc.conf..259R} \\ 
1999/03 & $31.32 \pm 0.15$ & $M_{V,{\rm TO}}^{\rm MW} = -7.61$ mag & $-0.11$ & 31.21 & NGC 1316 & & \citet{2000fepc.conf..259R} \\ 
1999/03 & $31.0  \pm 0.2 $ & $M_{V,{\rm TO}}^{\rm MW} = -7.4 $ mag & $+0.10$ & 31.10 & Fornax   & & \citet{1997AnA...319..470K} \\ 
1999/07 &  31.3            & $M_{V,{\rm TO}}^{\rm MW} = -7.4 $ mag & $+0.10$ & 31.40 & Fornax   & & \citet{1999AnAS..138...55H} \\ 
2000/02 & $31.38 \pm 0.15$ & $M_{V,{\rm TO}}^{\rm MW} = -7.60$ mag & $-0.10$ & 31.28 & Fornax   & $V$ & \citet{2000ApJ...529..745F} \\ 
2000/02 & $31.76 \pm 0.31$ & $M_{V,{\rm TO}}^{\rm MW} = -7.60$ mag & $-0.10$ & 31.66 & Fornax   & $B$ & \citet{2000ApJ...529..745F} \\ 
2001/06 & $31.58 \pm 0.18$ & $M_{V,{\rm TO}}^{\rm MW} = -7.60$ mag & $-0.10$ & 31.48 & NGC 1316 & $B$ & \citet{2001AnA...371..875G} \\ 
2001/06 & $31.47 \pm 0.22$ & $M_{V,{\rm TO}}^{\rm MW} = -7.60$ mag & $-0.10$ & 31.37 & NGC 1316 & $V$ & \citet{2001AnA...371..875G} \\ 
2001/06 & $31.19 \pm 0.17$ & $M_{V,{\rm TO}}^{\rm MW} = -7.60$ mag & $-0.10$ & 31.09 & NGC 1316 & $I$ & \citet{2001AnA...371..875G} \\ 
2005/01 &  31.4            & $M_{V,{\rm TO}}^{\rm MW} = -7.50$ mag & $ 0.00$ & 31.4  & NGC 1399 & & \citet{2005MNRAS.357...56F} \\ 
\hline 
\multicolumn{8}{c}{PNLF} \\
\hline
1990/09 &  30.88           & $(m-M)_0^{\rm M31} = 24.27$ mag & $+0.19$ & 31.07 & Fornax   & & \citet{1990BAAS...22.1312C} \\ 
1991/00 & $30.87 \pm 0.11$ & $(m-M)_0^{\rm M31} = 24.27$ mag & $+0.19$ & 31.06 & Fornax   & & cited by \citet{1993ApJ...415...10D} \\ 
1993/10 & $31.14 \pm 0.14$ & $(m-M)_0^{\rm M31} = 24.27$ mag & $+0.19$ & 31.33 & Fornax   & & \citet{1993ApJ...416...62M} \\ 
1993/10 & $31.13 \pm 0.06$ & $(m-M)_0^{\rm M31} = 24.27$ mag & $+0.19$ & 31.32 & NGC 1316 & & \citet{1993ApJ...416...62M} \\ 
1993/10 & $31.17 \pm 0.06$ & $(m-M)_0^{\rm M31} = 24.27$ mag & $+0.19$ & 31.36 & NGC 1399 & & \citet{1993ApJ...416...62M} \\ 
1993/12 & $31.19 \pm 0.07$ & $(m-M)_0^{\rm M31} = 24.32$ mag & $+0.14$ & 31.33 & NGC 1316 & & \citet{1993ApJ...419..479C} \\ 
1993/12 & $31.22 \pm 0.08$ & $(m-M)_0^{\rm M31} = 24.32$ mag & $+0.14$ & 31.36 & NGC 1399 & & \citet{1993ApJ...419..479C} \\ 
1996/05 & $31.24 \pm 0.06$ & $(m-M)_0^{\rm M31} = 24.24$ mag & $+0.19$ & 31.43 & Fornax   & & \citet{1997eds..proc..197J} \\ 
1999/03 & $31.33 \pm 0.08$ & $(m-M)_0^{\rm M31} = 24.44$ mag & $+0.02$ & 31.35 & Fornax   & & \citet{2000fepc.conf..259R} \\ 
1999/09 & $31.20 \pm 0.14$ & $+0.06$ mag w.r.t. \citet{1993ApJ...416...62M} & $+0.13$ & 31.33 & Fornax   & & \citet{1999AnARv...9..221L} \\ 
2000/02 & $31.20 \pm 0.07$ & $M^* = -4.58$ mag & $-0.10$ & 31.10 & Fornax   & & \citet{2000ApJ...529..745F} \\ 
2007/03 & $31.26 \pm 0.10$ & $M^* = -4.47$ mag & $+0.01$ & 31.27 & NGC 1316 & & \citet{2007ApJ...657...76F} \\ 
\hline 
\multicolumn{8}{c}{TFR} \\
\hline
1996/05 & $30.94 \pm 0.33$ & $(m-M)_0^{\rm Virgo} = 31.00$ mag & $+0.03$ & 30.97 & Fornax   & & \citet{1996ApJ...463...60B} \\ 
1999/03 & $31.43 \pm 0.12$ & $(m-M)_0^{\rm Virgo} = 31.58$ mag & $-0.55$ & 30.88 & Fornax   & & \citet{2000fepc.conf..259R} \\ 
1999/09 & $31.37 \pm 0.35$ & $(m-M)_0^{\rm Virgo} = 31.39$ mag & $-0.36$ & 31.01 & Fornax   & & \citet{1999AnARv...9..221L} \\ 
2000/04 & $31.25 \pm 0.50$ & $(m-M)_0^{\rm LMC}   = 18.50$ mag & $-0.01$ & 31.24 & Fornax   & & \citet{2000ApJ...533..744T} \\ 
2002/06 & $31.22 \pm 0.13$ & $(m-M)_0^{\rm LMC}   = 18.50$ mag & $-0.01$ & 31.21 & Fornax   & & \citet{2002AJ....123.2990B} \\ 
2004/01 & $31.57 \pm 0.33$ & $(m-M)_0^{\rm Virgo} = 31.40$ mag & $-0.37$ & 31.20 & Fornax   & & \citet{2004MNRAS.347.1011A} \\ 
2006/12 & $31.46 \pm 0.14$ & $(m-M)_0^{\rm LMC}   = 18.50$ mag & $-0.01$ & 31.45 & Fornax   & & \citet{2006ApJ...653..861M} \\ 
2009/08 & $31.38 \pm 0.06$ & $(m-M)_0^{\rm LMC}   = 18.50$ mag & $-0.01$ & 31.37 & NGC 1365 & & \citet{2009AJ....138..323T} \\ 
2012/04 & $31.19 \pm 0.12$ & $(m-M)_0^{\rm LMC}   = 18.50$ mag & $-0.01$ & 31.18 & Fornax   & & \citet{2012ApJ...749...78T} \\ 
2013/03 & $31.20 \pm 0.14$ & $(m-M)_0^{\rm LMC}   = 18.48$ mag & $+0.01$ & 31.21 & Fornax   & 3.6 $\mu$m & \citet{2013ApJ...765...94S} \\ 
2014/10 & $31.16 \pm 0.13$ & $(m-M)_0^{\rm LMC}   = 18.48$ mag & $+0.01$ & 31.17 & Fornax   & & \citet{2014MNRAS.444..527S} \\ 
\hline
\hline 
\end{tabular}
}
\end{center}
\flushleft 
$^a$ The calibration of our distance framework is based on distance
moduli to the LMC, M31, NGC 4258, and the Virgo cluster of
$(m-M)_0^{\rm LMC} = 18.49$ mag (Paper I), $(m-M)_0^{\rm M31} = 24.46$
mag (Paper II), $(m-M)_0^{\rm Virgo} = 31.06$ mag (Paper VI), and
$(m-M)_0^{\rm N4258} = 29.29$ mag \citep{1999Natur.400..539H}. In
addition, we have adopted $M_I^{\rm TRGB} = -4.05$ mag \citep[TRGB
  magnitude in the $I$ band;][]{2004AnA...424..199B}, $M^* = -4.67$
mag \citep[PNLF cut-off magnitude at 5007
  {\AA};][]{1989ApJ...339...53C}, $M_{V,{\rm TO}}^{\rm MW} = -7.50$
mag \citep[GCLF turnover magnitude in the Milky
  Way;][]{1996AJ....112.1487H}, and a Galactic Center distance $R_0 =
8.3$ kpc (paper IV). \\
$^b$ This calibration corresponds to $(m-M)_0^{\rm Virgo} = 31.03$ mag
(based on group membership) and $(m-M)_0^{\rm M31} = 24.44$ mag. We
adopted a mean adjustment of $+0.01$ mag to reconcile these
calibration choices with our distance framework.
\end{table*}

\begin{table*}
\vspace{2cm}
\caption{Corrections to published Coma cluster distance moduli.}
\begin{center}
\label{tab3}
\tabcolsep 0.5mm
{\tiny
\begin{tabular}{@{}cclccccl@{}}
\hline \hline
\multicolumn{1}{c}{Date} & \multicolumn{1}{c}{$(m-M)_0^{\rm orig}$} &
\multicolumn{1}{c}{Orig. calibration} & \multicolumn{1}{c}{Correction} &
\multicolumn{1}{c}{$(m-M)_0^{\rm scaled}$} &
\multicolumn{1}{c}{Target} & \multicolumn{1}{c}{Notes} &
\multicolumn{1}{c}{Reference} \\
\multicolumn{1}{c}{(yyyy/mm)} & \multicolumn{1}{c}{(mag)} & &
\multicolumn{1}{c}{(mag)} & \multicolumn{1}{c}{(mag)} \\
\hline 
\multicolumn{8}{c}{SBF} \\
\hline
1997/07 & $35.04 \pm 0.31$ & $(m-M)_0^{\rm M31}   = 24.43$ mag      & $+0.03$ & 35.07 & NGC 4881 & & \citet{1997ApJ...483L..37T} \\
1998/00 & $35.1  \pm 0.3 $ & $(m-M)_0^{\rm M31}   = 24.43$ mag      & $+0.03$ & 35.13 & NGC 4881 & & \citet{1998MmSAI..69..155B} \\
1999/01 & $34.64 \pm 0.25$ & $(m-M)_0^{\rm Virgo} = 31.06$ mag      & $-0.03$ & 34.61 & NGC 4889 & & \citet{1999ApJ...510...71J} \\
2000/02 & $35.10 \pm 0.55$ & $+0.05$ mag w.r.t. \citet{1997ApJ...475..399T} & $-0.02$ & 35.08 & Coma     & & \citet{2000ApJ...529..745F} \\
2001/05 & $35.05 \pm 0.50$ & $+0.05$ mag w.r.t. \citet{1997ApJ...475..399T} & $-0.02$ & 35.03 & NGC 4881 & & \citet{2001ApJ...553...47F} \\
2001/08 & $34.99 \pm 0.21$ & $+0.05$ mag w.r.t. \citet{1997ApJ...475..399T} & $-0.02$ & 34.97 & Coma     & $K$ & \citet{2001ApJ...557L..31L} \\
2013/02 &  35.             & $(m-M)_0^{\rm Virgo} = 31.09$ mag      & $-0.06$ & 34.94 & Coma     & & \citet{2013IAUS..289..304B} \\
\hline
\multicolumn{8}{c}{GCLF} \\
\hline
1995/12 & $35.16 \pm 0.23$ & $(m-M)_0^{\rm M31} = 24.6 $ mag       & $-0.14$ & 35.02 & NGC 4881 & lower limit & \citet{1995AJ....110.2537B}  \\
1996/12 & $35.15 \pm 0.06$ & $(m-M)_0^{\rm M31} = 24.6 $ mag       & $-0.14$ & 35.01 & IC 4051  & & \citet{1996AAS...189.1204B} \\
1996/12 & $35.07 \pm 0.17$ & $(m-M)_0^{\rm M31} = 24.6 $ mag       & $-0.14$ & 34.93 & IC 4051  & & \citet{1996AAS...189.1204B} \\
1997/05 & $35.15 \pm 0.16$ & $(m-M)_0^{\rm M31} = 24.6 $ mag       & $-0.14$ & 35.01 & IC 4051  & & \citet{1997AJ....113.1483B} \\
1997/05 & $35.07 \pm 0.17$ & $(m-M)_0^{\rm M31} = 24.6 $ mag       & $-0.14$ & 34.93 & IC 4051  & & \citet{1997AJ....113.1483B} \\
1997/05 & $35.11 \pm 0.12$ & $(m-M)_0^{\rm M31} = 24.6 $ mag       & $-0.14$ & 34.97 & IC 4051  & & \citet{1997AJ....113.1483B} \\
1998/07 & $35.13 \pm 0.15$ & $(m-M)_0^{\rm M31} = 24.6 $ mag       & $-0.14$ & 34.99 & IC 4051  & & \citet{1998AJ....116...31B} \\
1998/07 & $35.38 \pm 0.16$ & $(m-M)_0^{\rm M31} = 24.6 $ mag       & $-0.14$ & 35.24 & IC 4051  & & \citet{1998AJ....116...31B} \\
1999/00 & $34.34 \pm 0.31$ & $M_{V,{\rm TO}}^{\rm MW} = -7.62$ mag & $-0.12$ & 34.22 & Coma     & & \citet{1999ASPC..167..204T} \\
2000/04 & $35.05 \pm 0.12$ & $M_{V,{\rm TO}}^{\rm MW} = -7.26$ mag & $+0.14$ & 35.19 & Coma     & & \citet{2000ApJ...533..125K} \\
2011/00 &  34.9            & $M_{V,{\rm TO}}^{\rm MW} = -7.5 $ mag & $ 0.00$ & 34.9  & NGC 4921 & & \citet{2011AstL...37..766T} \\
2011/00 &  35.0            & $M_{V,{\rm TO}}^{\rm MW} = -7.5 $ mag & $ 0.00$ & 35.0  & NGC 4923 & & \citet{2011AstL...37..766T} \\
2011/00 & $34.93 \pm 0.11$ & $M_{V,{\rm TO}}^{\rm MW} = -7.5 $ mag & $ 0.00$ & 34.93 & Coma     & & \citet{2011AstL...37..766T} \\
2016/03 & $34.80 \pm 0.09$ & $M_{V,{\rm TO}}^{\rm MW} = -7.66$ mag & $-0.16$ & 34.64 & Coma     & & \citet{2016ApJ...819...77L} \\
2016/03 &  34.65           & $M_{V,{\rm TO}}^{\rm MW} = -7.66$ mag & $-0.16$ & 34.49 & Coma     & & \citet{2016ApJ...819...77L} \\
\hline
\multicolumn{8}{c}{TFR} \\
\hline
1991/07 & $34.5  \pm 0.4 $ & $(m-M)_0^{\rm M31} = 24.2 $ mag & $+0.26$ & 34.76 & Coma     & $B$ & \citet{1991ApJ...376....8F} \\
1991/07 & $34.37 \pm 0.1 $ & $(m-M)_0^{\rm M31} = 24.2 $ mag & $+0.26$ & 34.63 & Coma     & $B$ & \citet{1991ApJ...376....8F} \\
1991/07 &  33.9            & $(m-M)_0^{\rm M31} = 24.2 $ mag & $+0.26$ & 34.16 & Coma     & $B$ & \citet{1991ApJ...376....8F} \\
1991/07 &  34.5            & $(m-M)_0^{\rm M31} = 24.2 $ mag & $+0.26$ & 34.76 & Coma     & $B$ & \citet{1991ApJ...376....8F} \\
1993/07 & $34.32 \pm 0.07$ & $(m-M)_0^{\rm M31} = 24.37$ mag & $+0.09$ & 34.41 & Coma     & $B$ & \citet{1993MNRAS.263..211R} \\
1993/07 & $34.56 \pm 0.16$ & $(m-M)_0^{\rm M31} = 24.37$ mag & $+0.09$ & 34.65 & Coma     & $H$ & \citet{1993MNRAS.263..211R} \\
1994/06 & $34.86 \pm 0.14$ & $(m-M)_0^{\rm M31} = 24.43$ mag & $+0.03$ & 34.89 & Coma     & & \citet{1994AJ....107.1962B} \\
1994/06 & $34.54 \pm 0.22$ & $(m-M)_0^{\rm M31} = 24.43$ mag & $+0.03$ & 34.57 & Coma     & & \citet{1994AJ....107.1962B} \\
1994/06 & $33.90 \pm 0.32$ & $(m-M)_0^{\rm M31} = 24.43$ mag & $+0.03$ & 33.93 & Coma     & & \citet{1994AJ....107.1962B} \\
1998/00 &  34.67           & $(m-M)_0^{\rm LMC} = 18.50$ mag & $-0.01$ & 34.66 & Coma     & $BRI$ & \citet{1998MmSAI..69..237T} \\
2000/04 & $34.68 \pm 0.15$ & $(m-M)_0^{\rm LMC} = 18.50$ mag & $-0.01$ & 34.67 & Coma     & $I$ & \citet{2000ApJ...533..744T} \\
2001/05 &  34.66           & $(m-M)_0^{\rm LMC} = 18.50$ mag & $-0.01$ & 34.65 & Coma     & $I$ & \citet{2001ApJ...553...47F} \\
2001/07 & $34.94 \pm 0.13$ & $(m-M)_0^{\rm M31} = 24.44$ mag & $+0.02$ & 34.96 & Coma     & IR & \citet{2001ApJ...555..215W} \\
2004/05 & $34.58 \pm 0.30$ & $(m-M)_0^{\rm M31} = 24.48$ mag & $-0.02$ & 34.56 & Coma     & & \citet{2004ApJ...607..241R} \\
2004/05 & $34.76 \pm 0.33$ & $(m-M)_0^{\rm M31} = 24.48$ mag & $-0.02$ & 34.74 & Coma     & & \citet{2004ApJ...607..241R} \\
2007/10 & $34.80 \pm 0.11$ & $(m-M)_0^{\rm LMC} = 18.50$ mag & $-0.01$ & 34.79 & Coma     & & \citet{2007ApJS..172..599S} \\
2009/08 &  34.65           & $(m-M)_0^{\rm LMC} = 18.39$ mag & $+0.11$ & 34.76 & NGC 4881 & $K$ & \citet{2009JApA...30...93R} \\
2009/08 &  34.59           & $(m-M)_0^{\rm LMC} = 18.39$ mag & $+0.11$ & 34.70 & NGC 4881 & $K$ & \citet{2009JApA...30...93R} \\
2009/08 &  34.70           & $(m-M)_0^{\rm LMC} = 18.39$ mag & $+0.11$ & 34.81 & Coma     & $K$ & \citet{2009JApA...30...93R} \\
2009/08 &  34.77           & $(m-M)_0^{\rm LMC} = 18.39$ mag & $+0.11$ & 34.88 & Coma     & $K$ & \citet{2009JApA...30...93R} \\
2011/06 & $34.40 \pm 0.08$ & $M_I^{\rm TRGB}    = -4.05$ mag & $ 0.00$ & 34.40 & Coma     & & \citet{2011ApJ...733...75H} \\
2012/04 & $34.77 \pm 0.10$ & $(m-M)_0^{\rm LMC} = 18.50$ mag & $-0.01$ & 34.76 & Coma     & $I$ & \citet{2012ApJ...749...78T} \\
2012/04 & $34.83 \pm 0.06$ & $(m-M)_0^{\rm LMC} = 18.50$ mag & $-0.01$ & 34.82 & Coma     & & \citet{2012ApJ...749..174C} \\
2012/10 & $34.90 \pm 0.13$ & $(m-M)_0^{\rm LMC} = 18.48$ mag & $+0.01$ & 34.91 & Coma     & 3.6 $\mu$m & \citet{2012ApJ...758L..12S} \\
2013/03 & $34.90 \pm 0.13$ & $(m-M)_0^{\rm LMC} = 18.48$ mag & $+0.01$ & 34.91 & Coma     & mid-IR & \citet{2013ApJ...765...94S} \\
2013/10 & $35.10 \pm 0.22$ & $(m-M)_0^{\rm LMC} = 18.48$ mag & $+0.01$ & 35.11 & NGC 4889 & $I$ & \citet{2013AJ....146...86T} \\
2013/10 & $34.82 \pm 0.22$ & $(m-M)_0^{\rm LMC} = 18.48$ mag & $+0.01$ & 34.83 & NGC 4874 & $I$ & \citet{2013AJ....146...86T} \\
2014/09 & $34.91 \pm 0.06$ & $(m-M)_0^{\rm LMC} = 18.48$ mag & $+0.01$ & 34.92 & Coma     & $W1$ & \citet{2014ApJ...792..129N} \\
2014/09 & $34.94 \pm 0.06$ & $(m-M)_0^{\rm LMC} = 18.48$ mag & $+0.01$ & 34.95 & Coma     & $W2$ & \citet{2014ApJ...792..129N} \\
2014/09 & $34.86 \pm 0.11$ & $(m-M)_0^{\rm LMC} = 18.48$ mag & $+0.01$ & 34.87 & Coma     & $W1$ & \citet{2014ApJ...792..129N} \\
2014/09 & $34.87 \pm 0.11$ & $(m-M)_0^{\rm LMC} = 18.48$ mag & $+0.01$ & 34.88 & Coma     & $W2$ & \citet{2014ApJ...792..129N} \\
2014/09 & $34.77 \pm 0.10$ & $(m-M)_0^{\rm LMC} = 18.48$ mag & $+0.01$ & 34.78 & Coma     & $I$ & \citet{2014ApJ...792..129N} \\
2014/10 & $34.78 \pm 0.11$ & $(m-M)_0^{\rm LMC} = 18.48$ mag & $+0.01$ & 34.79 & Coma     & & \citet{2014MNRAS.444..527S} \\
\hline
\hline 
\end{tabular}
}
\end{center}
\end{table*}

\addtocounter{table}{-1}
\begin{table*}
\vspace{2cm}
\caption{(Continued)}
\begin{center}
\tabcolsep 0.5mm
{\tiny
\begin{tabular}{@{}cclccccl@{}}
\hline \hline
\multicolumn{1}{c}{Date} & \multicolumn{1}{c}{$(m-M)_0^{\rm orig}$} &
\multicolumn{1}{c}{Orig. calibration} & \multicolumn{1}{c}{Correction$^a$} &
\multicolumn{1}{c}{$(m-M)_0^{\rm scaled}$} &
\multicolumn{1}{c}{Target} & \multicolumn{1}{c}{Notes} &
\multicolumn{1}{c}{Reference} \\
\multicolumn{1}{c}{(yyyy/mm)} & \multicolumn{1}{c}{(mag)} & &
\multicolumn{1}{c}{(mag)} & \multicolumn{1}{c}{(mag)} \\
\hline
\multicolumn{8}{c}{Hubble law} \\
\hline
1998/06 &  34.21           & $H_0 = 100 \, {\rm km \, s}^{-1} \, {\rm Mpc}^{-1}$ & $+0.77$ & 34.98 & Coma & & \citet{1998ApJ...500..750K} \\
1999/12 &  34.2            & $H_0 = 100 \, {\rm km \, s}^{-1} \, {\rm Mpc}^{-1}$ & $+0.77$ & 34.97 & Coma & & \citet{1999MNRAS.310..445T} \\
2002/03 &  35.13           & $H_0 =  70 \, {\rm km \, s}^{-1} \, {\rm Mpc}^{-1}$ & $ 0.00$ & 35.13 & Coma & & \citet{2002ApJ...567..130B} \\
2003/08 &  35.03           & $H_0 =  70 \, {\rm km \, s}^{-1} \, {\rm Mpc}^{-1}$ & $ 0.00$ & 35.03 & Coma & & \citet{2003MNRAS.343..401L} \\
2008/06 &  34.98           & $H_0 =  71 \, {\rm km \, s}^{-1} \, {\rm Mpc}^{-1}$ & $+0.03$ & 35.01 & Coma & & \citet{2008ApJS..176..424C} \\
2009/02 & $34.97 \pm 0.13$ & $H_0 =  72 \, {\rm km \, s}^{-1} \, {\rm Mpc}^{-1}$ & $+0.06$ & 35.03 & Coma & & \citet{2009AJ....137.3314H} \\
2015/01 &  34.96           & $H_0 =  70 \, {\rm km \, s}^{-1} \, {\rm Mpc}^{-1}$ & $ 0.00$ & 34.96 & Coma & & \citet{2015ApJ...798L..45V} \\
2015/01 &  35.06           & $H_0 =  70 \, {\rm km \, s}^{-1} \, {\rm Mpc}^{-1}$ & $ 0.00$ & 35.06 & Coma & & \citet{2015ApJ...798L..45V} \\
\hline
\hline 
\end{tabular}
}
\end{center}
\end{table*}

From a sociological perspective, we note that distance measures using
a specific tracer are often dominated by articles published by the
same group and their junior team members. This is not unexpected, of
course, since this practice reflects the central expertise of the
different groups of authors. Nevertheless, we considered the effects
of including series of results from the same group on the overall
value of the resulting distance modulus.

\subsection{Fornax}

First, we considered the post-1990 Cepheid-based distance moduli for
the Fornax cluster. Comparing author lists, combined with a careful
perusal of the papers in question, it is clear that among our set of
13 Cepheid distances to the Fornax cluster, the only truly independent
measurement was provided by \citet{2016ApJ...826...56R}, i.e.,
$(m-M)_0 = 31.21 \pm 0.06$ mag (after adjustment of its zero-point
calibration). Their value falls within the mutual 1$\sigma$
uncertainties of the Cepheid-based distance included in Table
\ref{tab1}; one should, of course, keep in mind that all other
measurements contributing to that distance are correlated and not
independent.

Among the eight post-2000 TFR-based distances to the Fornax cluster,
five were published by the same group. The weighted average of those
five determinations is $(m-M)_0 = 31.22 \pm 0.25$ mag. Of the
remaining three values, two fall comfortably within the 1$\sigma$
uncertainties following homogenization: $(m-M)_0 = 31.21 \pm 0.13$ mag
\citep{2002AJ....123.2990B} and $(m-M)_0 = 31.20 \pm 0.33$ mag
\citep{2004MNRAS.347.1011A}. The third value, $(m-M)_0 = 31.45 \pm
0.14$ mag \citep{2006ApJ...653..861M}, is larger although still
consistent with the bulk of our values.

Nevertheless, inspection of Table \ref{tab1} shows that the TFR-based
weighted mean distance to the Fornax cluster is significantly shorter
than the equivalent distance estimates based on most other tracers
(with the notable exception of GCLF-based distance moduli; see
below). A combination of effects may have given rise to this
difference. First, \citet{2004MNRAS.347.1011A} concluded that for
their sample of 18 galaxies at distances of $(m-M)_0 > 29.5$ mag
hosting Cepheid variables and which were observed with the {\sl Hubble
  Space Telescope}, the TFR-based distance moduli yield distances that
are shorter by $\Delta (m-M)_0 = -0.44 \pm 0.09$ mag (see also Shanks
1997). They suggested that at least some of this effect may be owing
to unaccounted-for metallicity differences
\citep[e.g.][]{1998ApJ...498..181K} and sample incompleteness, leading
to a significant scale error in TFR-based distances. We note, however,
that addition of the offset in distance modulus suggested by Shanks
(1997) and \citet{2004MNRAS.347.1011A} would lead to overestimated TFR
distances compared with Fornax distance moduli based on other
tracers. This situation is exacerbated by the often convoluted
calibration approaches, often involving at least some iterative means
to tie Cepheid, Type Ia supernovae (SNe Ia), and TFR distances to the
same scale.

In addition, TFR distances tend to differ depending on the operating
wavelength. \citet{2000ApJ...533..744T} found for the Ursa Major
galaxy cluster that although the overall agreement among the distance
moduli resulting from analysis of different passbands is good, their
$I$-band analysis yielded shorter distance moduli than the weighted
mean by 0.02 mag, while in the $B$ band their moduli were
overestimated by 0.04 mag. In the $R$ and $K'$ bands, their estimates
were 0.03 mag larger and 0.05 mag smaller, respectively, than the
mean. Finally, a degree of publication bias could have crept into our
sample of TFR-based Fornax cluster distance measures, given that some
authors confidently state that their derived TFR-based distance moduli
comfortably agree with previously published measures, but without
comparing the underlying calibrations applied
\citep[e.g.][]{1996ApJ...463...60B}.

Our SBF-based distance measures to Fornax represent the largest
subsample. Although they do exhibit some spread about the weighted
mean, the distribution's standard deviation is small, 0.14 mag, and
therefore not indicative of statistical anomalies. Nevertheless, 21 of
the 28 post-2000 values considered here were published by the same
team (and are, hence, likely correlated). However, the remaing seven
values \citep{2000ApJ...529..745F, 2002ApJ...564..216L,
  2003AnA...398...63J, 2006AJ....132.1384D} are all fully consistent
with the overall weighted mean and its 1$\sigma$ uncertainty. Both
data sets are statistically indistinguishable. We note that our three
TRGB-based Fornax distances were all published by the same team, but
we included them in our analysis because they provide an independent
stellar population tracer. Both the SBF technique and the TRGB-based
distances rely on red-giant stars; it is therefore comforting to see
that the distance moduli resulting from independent application of
these techniques are indeed very close to one another.

The GCLF-based data set comprises 19 distance moduli. The author lists
of the contributing papers are more diverse than for the previously
discussed tracers. The largest single group of collaborators
contributed to nine of the measures included in our catalog. Overall,
assuming a Gaussian distribution of distance measures, the 1$\sigma$
spread is 0.19 mag, similar to the uncertainty on the mean. This
indicates that the intrinsic spread among the contributing values is
more significant than the equivalent spreads resulting from the other
tracers used here. Some of the most significantly deviating values
result from calibrations using non-standard calibrators, specifically
the poorly defined $B$- or $I$-band GCLFs pertaining to the Milky Way
or M31 \citep[e.g.][]{1996AnA...309L..39K, 2000ApJ...529..745F,
  2001AnA...371..875G}. In addition, calibration of the ($V$-band)
GCLF in the Milky Way relies on accurate distance determinations to
the contributing globular clusters (and, in fact, a reliable distance
determination to the Galactic Center; see Paper IV). Moreover, one
must make assumptions regarding the shape (width) of the GCLF, which
may differ among different galaxy types
\citep[e.g.][]{2006ApJ...651L..25J}.

Finally, as we discussed in the context of the distance to the
Galactic Center (Paper IV), distances based on GCLFs tend to be
systematically smaller than most other distance measurements (for the
Fornax cluster, see also Villegas et al. 2010). This could be caused
by incomplete corrections for internal or foreground extinction, or
because of incomplete samples of objects, in the sense that our
observational data may be biased toward objects in the foreground of
the host galaxy. In view of these lingering uncertainties, we will
refrain from further consideration of the GCLF-based distances.

The five articles yielding PNLF-based distances to Fornax were
published by four different groups, yet following homegenization the
weighted mean is well-defined with a small uncertainty (0.09
mag). Nevertheless, we are reluctant to place too much emphasis on
this result, given that -- like the GCLF -- planetary nebulae samples
are often dominated by objects located predominantly in the foreground
of their host galaxies, thus resulting in underestimated distances
(for a discussion, see Paper VI).

In summary, we argue that the most reliable Fornax distance moduli
among the values in our database are those resulting from analyses of
Cepheid distances, SBF, and the TRGB. Their weighted mean results in
\begin{eqnarray}
(m-M)_0^{\rm Fornax} &=& 31.41 \pm 0.15 \mbox{ mag} \nonumber \\
\mbox{or } D &=& 19.1^{+1.4}_{-1.2} \mbox{ Mpc}. \nonumber
\end{eqnarray}

For completeness, we also considered the post-1990 distance moduli
that were not included in our analysis, including those based on SNe
Ia, novae, application of the Hubble law, and other, less commonly
used tracers (see the online table at
http://astro-expat.info/Data/pubbias.html). With few exceptions, the
vast majority of these post-1990 measures were comfortably consistent
with the weighted mean distance modulus derived above.

\subsection{Coma}

We will now biefly review the published Coma cluster distance moduli
along the same lines as we just did for the Fornax cluster. Table
\ref{tab1} includes four different distance measures. As we argued
above for the Fornax cluster, use of GCLF-based distances is fraught
with lingering uncertainties, and so we will not consider those
measurements here. Of the remaining three tracers, use of the Hubble
law requires a somewhat different analysis. The Coma cluster distance
coincides with the distance where the `smooth' Hubble flow starts,
i.e., where redshifts of field galaxies become reasonably reliable
proxies of their distances. \citet{2009ApJ...699..539R} suggested a
minimum redshift of $z = 0.023$ ($D \sim 100$ Mpc) for the smooth
Hubble flow. At $D = 100$ Mpc, the Hubble-flow velocity is around 7000
km s$^{-1}$ and peculiar velocities will typically amount to a 5\%
contribution.

The main uncertainty in this context relates to the value of the
Hubble parameter; radial velocities to individual galaxies and even to
entire galaxy clusters can be determined to high accuracy and
precision. The mean distance modulus to the Coma cluster based on its
recession velocity has been homogenized to a Hubble parameter of $H_0
= 70$ km s$^{-1}$ Mpc$^{-1}$, adopted as compromise value given the
prevailing 1--2$\sigma$ tension remaining in this field
\citep[e.g.][and references therein]{2019ApJ...876...85R}. A reduction
(increase) to $H_0 = 67.3$ (72) km s$^{-1}$ Mpc$^{-1}$ would result in
an increase (decrease) in the Coma distance modulus by 0.09 (0.06)
mag. The `uncertainty' associated with this method included in the
table relfects the Gaussian $\sigma$ of the distribution (since onle
one of the measurements included in our final catalog quoted
uncertainties). Using the single uncertainty estimate published in
this context \citep{2009AJ....137.3314H}, a more realistic uncertainty
would require addition in quadrature of this 0.13 mag uncertainty,
resulting in a total error of order 0.14 mag.

As for the Fornax cluster, the TFR-based distances to the Coma cluster
are systematically shorter than our other distances. We will therefore
not include TFR-based distance measures in our analysis. This thus
leaves the SBF-based Coma cluster distances. Four of the seven values
relate to the SBF distance to NGC 4881, with the remaining three
referring to the Coma cluster as a whole. None of the groups
contributing to its weighted mean dominate the set of values, so we
have no reason to suspect correlated measurements.

In summary, if we adopt the Hubble law- and SBF-based distance moduli
(with our updated uncertainty for the Hubble distances), we find a
Coma cluster distance modulus of
\begin{eqnarray}
(m-M)_0^{\rm Coma} &=& 34.99 \pm 0.38 \mbox{ mag} \nonumber \\ 
\mbox{or } D &=& 99.5^{+19.0}_{-15.9} \mbox{ Mpc}. \nonumber
\end{eqnarray}

For completeness, we again considered the post-1990 distance moduli
that were not included in our analysis, including those based on SNe
Ia, Fundamental Plane scaling, and other, less commonly used tracers
(see the online table at
http://astro-expat.info/Data/pubbias.html). All of the latter
post-1990 measures were comfortably consistent with the weighted mean
distance modulus derived above.

\subsection{Relative distance moduli}
\label{relative.sec}

At distances equivalent to or beyond that of the Virgo cluster, it has
become relatively common to quote distance measures relative to the
Virgo cluster. The main advantage of using relative rather than
absolute distances is that the number of assumptions one has to make
is significantly reduced, provided that the physical basis on which
the distances are compared is similar. For the Fornax cluster, our
database includes 43 relative Fornax--Virgo cluster distance moduli
published since 1990. A straight weighted mean yields $\Delta
(m-M)_0^{\rm Fornax-Virgo} = 0.18 \pm 0.28$ mag. We used all 43 values
to determine the central value of this relative distance modulus;
since five measurements do not have any uncertainties associated with
them, we used the remaining 38 values to determine the uncertainty
pertaining to our weighted mean. Both clusters thus appear to be
located at very similar distance, with Fornax placed marginally more
distant.

To avoid any effects associated with small-number statistics, we
subdivided our 43 relative distance moduli into three groups of
similar tracers, yielding post-1990 weighted-mean relative distance
moduli of $\Delta (m-M)_0^{\rm Fornax-Virgo} = 0.23 \pm 0.18$ mag for
the 11 SBF-based measurements, $\Delta (m-M)_0^{\rm Fornax-Virgo} =
0.10 \pm 0.12$ mag for the 10 values based on luminosity functions
(GCLF, PNLF), and $\Delta (m-M)_0^{\rm Fornax-Virgo} = 0.15 \pm 0.15$
mag for the 11 relative distance measures based on kinematic tracers
(TFR, the $D_n-\sigma$ relation, and Fundamental Plane
scaling). Combined with our recommended Virgo cluster distance modulus
of $(m-M)_0^{\rm Virgo} = 31.03 \pm 0.14$ mag (Paper VI), these
relative measurements convert to absolute Fornax cluster distance
moduli of
\begin{itemize}
\item $(m-M)_0^{\rm Fornax} = 31.21 \pm 0.28$ mag (all values);
\item $(m-M)_0^{\rm Fornax,SBF} = 31.26 \pm 0.23$ mag (SBF-based measures);
\item $(m-M)_0^{\rm Fornax,LFs} = 31.13 \pm 0.12$ mag (luminosity
  functions); and
\item $(m-M)_0^{\rm Fornax,kin} = 31.18 \pm 0.15$ mag (kinematic tracers);
\end{itemize}
The weighted mean of the latter three values is
\begin{eqnarray}
(m-M)_0^{\rm Fornax,stat} &=& 31.18 \pm 0.17 \mbox{ mag} \nonumber \\
\mbox{or } D &=& 17.2^{+1.4}_{-1.3} \mbox{ Mpc}. \nonumber
\end{eqnarray}

As for the Coma cluster, relative distance moduli have been determined
with respect to the Virgo and Fornax clusters and the Leo I group. The
23 post-1990 relative Coma--Virgo distance moduli (of which 22 have
associated uncertainties) yield a straight weighted mean of $\Delta
(m-M)_0^{\rm Coma-Virgo} = 3.75 \pm 0.23$ mag. The subset of six
kinematics-based relative distance measures (TFR, Fundamental Plane)
result in $\Delta (m-M)_0^{\rm Coma-Virgo} = 3.63 \pm 0.22$ mag. (None
of the other possible subsets of Coma--Virgo relative distance
determinations reach the threshold where a statistical analysis
becomes meaningful.)

Combining the weighted mean of the full set of measurements with our
preferred Virgo cluster distance modulus yields
\begin{eqnarray}
(m-M)_0^{\rm Coma} &=& 34.78 \pm 0.27 \mbox{ mag} \nonumber \\
\mbox{or } D &=& 90.4^{+11.9}_{-10.6} \mbox{ Mpc}, \nonumber
\end{eqnarray}
while the kinematics-based distances yield
\begin{eqnarray}
(m-M)_0^{\rm Coma} &=& 34.66 \pm 0.26 \mbox{ mag} \nonumber \\
\mbox{or } D &=& 85.5^{+10.9}_{-9.6} \mbox{ Mpc}. \nonumber
\end{eqnarray}

Second, a significant number of relative distance moduli with respect
to the Leo I group were published by \citet{1997NewA....1..363G}, in
the form of Coma distance moduli based on a variety of independent
calibrations. His unweighted mean distance modulus to the Coma
cluster, $(m-M)_0 = 34.90 \pm 0.13$ mag, is based on a mean distance
ratio of $D_{\rm Coma} / D_{\rm Leo\, I} = 8.84 \pm 0.23$,
corresponding to $\Delta (m-M)_0^{\rm Coma-Leo\, I} = 4.73 \pm 0.06$
mag. However, on closer inspection, the data set underlying these
values raises a number of concerns. If one calculates the individual
Coma/Leo I distance ratios using the values included in his Table 2,
the central value of the resulting ratio is $D_{\rm Coma} / D_{\rm
  Leo\, I} = 8.84$ for every single, presumably independent
calibration method.

In the preamble to his Section 3, \citet{1997NewA....1..363G} states
that he derived the Leo I distances included in his Table 2 on the
basis of four different calibration methods. In addition, he states
that the Coma cluster distances included are based on the zero-point
calibration of the distance--velocity dispersion relation for Coma. If
the Leo I calibration methods applied were indeed independent (which
we have no reason to doubt), this very tightly defined central value
is statistically highly unlikely. Therefore, we decided to discard
\citet{1997NewA....1..363G}'s measurements, since we cannot ascertain
their integrity. This has unintended consequences, however, because a
number of subsequently published Coma distance moduli were also based
on this result \citep{1998MmSAI..69..127C, 1998MNRAS.298..166S}, and
so we were forced to discard them.

\citet{1997ApJ...483L..37T} provided the only independent relative
distance modulus between the Coma cluster and the Leo I group, $\Delta
(m-M)_0^{\rm Coma-Leo\, I} = 4.89 \pm 0.30$ mag. Although this
measurement is consistent, within the 1$\sigma$ uncertainties, with
the value promoted by \citet{1997NewA....1..363G}, we will
nevertheless refrain from further analysis of the Coma--Leo I distance
differential.

Finally, \citet{1994PASP..106.1113V} cited a distance ratio of $D_{\rm
  Coma}/D_{\rm Fornax} = 5.25 \pm 0.38$, although without providing
provenance. This corresponds to a relative distance modulus between
the Coma and Fornax clusters of $\Delta (m-M)_0^{\rm Coma-Fornax} =
3.60 \pm 0.15$ mag. Combining this with the Fornax cluster distance
moduli obtained above, we obtain
\begin{eqnarray}
(m-M)_0^{\rm Coma} &=& 34.78 \pm 0.23 \mbox{ mag}, \nonumber \\
                   &=& 34.81 \pm 0.32 \mbox{ mag, and} \nonumber \\
                   &=& 35.01 \pm 0.21 \mbox{ mag} \nonumber
\end{eqnarray}
for a subset of relative Coma--Virgo distance moduli, the full set of
Coma--Virgo measures, and our best direct estimate for the Virgo
cluster distance, respectively.

\section{A distance framework out to 100 Mpc}
\label{summary.sec}

In this paper, we have considered the body of published distance
moduli to the Fornax and Coma clusters, with specific emphasis on the
period since 1990. We carefully homogenized our final catalogs of
distance moduli onto the distance scale established in Papers I
through VI. We assessed systematic differences associated with the use
of specific tracers, and consequently discarded results based on
application of the TFR and of luminosity functions.

We recommend `best' weighted relative distance moduli for the Fornax
and Coma clusters with respect to the Virgo cluster benchmark of
$\Delta (m-M)_0^{\rm Fornax - Virgo} = 0.18 \pm 0.28 $ mag and $\Delta
(m-M)_0^{\rm Coma - Virgo} = 3.75 \pm 0.23$ mag. On balance, the set
of weighted mean distance moduli we derived as most representative of
the clusters' distances is as follows:
\begin{eqnarray}
(m-M)_0^{\rm Fornax} &=& 31.41 \pm 0.15 \mbox{ mag and} \nonumber \\
                     &=& 31.21 \pm 0.28 \mbox{ mag;} \nonumber \\
(m-M)_0^{\rm Coma}   &=& 34.99 \pm 0.38 \mbox{ mag and} \nonumber \\
                     &=& 34.78 \pm 0.27 \mbox{ mag}. \nonumber
\end{eqnarray}
For each cluster, this first distance modulus is the result of our
analysis of the direct, absolute distance moduli published since 1990,
while the second modulus is based on the relative measures published
during the same period. 

Interestingly, while the absolute and relative distance moduli for
both clusters are mutually consistent within the uncertainties, the
relative distance moduli yield absolute distances that are shorter by
$\sim 0.20$ mag, or about 1$\sigma$. It is unclear what may have
caused this small difference for both clusters; investigation of the
cause is beyond the scope of the present paper since it requires
careful examination of the individual distances comprising the tracers
commonly used in this field. It is unlikely that line-of-sight depth
effects are to blame \citep[e.g.][]{2003AnA...398...63J,
  2006AJ....132.1384D, 2009ApJ...694..556B}, given that most
individual galaxies and galaxy samples in both of our clusters
comprise the same or similar objects. We speculate that lingering
uncertainties in the underlying absolute distance scale appear to have
given rise to a systematic uncertainty of order 0.20 mag.

This concludes our series of papers aimed at establishing a robust and
internally consistent, statistically validated distance framework out
to distances of order 100 Mpc. Our recommended distances to the Fornax
and Coma clusters quoted above should be read in tandem with the
distance moduli we derived and recommend for the Galactic Center
(Paper IV), the Magellanic Clouds (Papers I and III), the M31 group
(Paper II), NGC 4258 \citep{1999Natur.400..539H}, and the Virgo cluster
(Paper VI): see Table 3 in Paper VI for the full set of recommended
distance moduli.

\section*{Acknowledgements}

This research has made extensive use of NASA's Astrophysics Data
System Abstract Service.


\begin{thebibliography}{}

\bibitem[Ajhar et al.(2001)]{2001ApJ...559..584A} Ajhar, E.~A., Tonry,
  J.~L., Blakeslee, J.~P., et al.\ 2001, \apj, 559, 584

\bibitem[Allen \& Shanks(2004)]{2004MNRAS.347.1011A} Allen, P.~D., \&
  Shanks, T.\ 2004, \mnras, 347, 1011

\bibitem[Baum(1998)]{1998AJ....116...31B} Baum, W.~A.\ 1998, \aj, 116,
  31

\bibitem[Baum et al.(1995)]{1995AJ....110.2537B} Baum, W.~A.,
  Hammergren, M., Groth, E.~J., et al.\ 1995, \aj, 110, 2537

\bibitem[Baum et al.(1996)]{1996AAS...189.1204B} Baum, W.~A.,
  Hammergren, M., Thomsen, B., et al.\ 1996, Am. Astron.  Soc. Mtg
  Abstr. 189, 12.04

\bibitem[Baum et al.(1997)]{1997AJ....113.1483B} Baum, W.~A.,
  Hammergren, M., Thomsen, B., et al.\ 1997, \aj, 113, 1483

\bibitem[Bellazzini et al.(2004)]{2004AnA...424..199B} Bellazzini, M.,
  Ferraro, F.~R., Sollima, A., et al.\ 2004, \aap, 424, 199

\bibitem[Bernardi et al.(2002)]{2002AJ....123.2990B} Bernardi, M.,
  Alonso, M.~V., da Costa, L.~N., et al.\ 2002, \aj, 123, 2990

\bibitem[Bernstein et al.(1994)]{1994AJ....107.1962B} Bernstein,
  G.~M., Guhathakurta, P., Raychaudhury, S., et al.\ 1994, \aj, 107,
  1962

\bibitem[Blakeslee(2013)]{2013IAUS..289..304B} Blakeslee, J.~P.\ 2013,
  in Advancing the Physics of Cosmic Distances, ed. R. de Grijs, IAU
  Symp. 289, 304

\bibitem[Blakeslee et al.(2010)]{2010ApJ...724..657B} Blakeslee,
  J.~P., Cantiello, M., Mei, S., et al.\ 2010, \apj, 724, 657

\bibitem[Blakeslee et al.(2009)]{2009ApJ...694..556B} Blakeslee,
  J.~P., Jord{\'a}n, A., Mei, S., et al.\ 2009, \apj, 694, 556

\bibitem[Blakeslee et al.(2001)]{2001MNRAS.320..193B} Blakeslee,
  J.~P., Vazdekis, A., \& Ajhar, E.~A.\ 2001, \mnras, 320, 193

\bibitem[Bothun et al.(1991)]{1991ApJ...376..404B} Bothun, G.~D.,
  Impey, C.~D., \& Malin, D.~F.\ 1991, \apj, 376, 404

\bibitem[Bridges et al.(1991)]{1991AJ....101..469B} Bridges, T.~J.,
  Hanes, D.~A., \& Harris, W.~E.\ 1991, \aj, 101, 469

\bibitem[Brighenti \& Mathews(2002)]{2002ApJ...567..130B} Brighenti,
  F., \& Mathews, W.~G.\ 2002, \apj, 567, 130

\bibitem[Brocato et al.(1998)]{1998MmSAI..69..155B} Brocato, E.,
  Capaccioli, M., \& Condelli, M.\ 1998, \memsai, 69, 155

\bibitem[Bureau et al.(1996)]{1996ApJ...463...60B} Bureau, M., Mould,
  J.~R., \& Staveley-Smith, L.\ 1996, \apj, 463, 60

\bibitem[Cantiello et al.(2007)]{2007ApJ...668..130C} Cantiello, M.,
  Blakeslee, J., Raimondo, G., et al.\ 2007, \apj, 668, 130

\bibitem[Cantiello et al.(2013)]{2013AnA...552A.106C} Cantiello, M.,
  Grado, A., Blakeslee, J.~P., et al.\ 2013, \aap, 552, A106

\bibitem[Carter et al.(2008)]{2008ApJS..176..424C} Carter, D.,
  Goudfrooij, P., Mobasher, B., et al.\ 2008, \apjs, 176, 424

\bibitem[Cassisi \& Salaris(1998)]{1998MmSAI..69..127C} Cassisi, S.,
  \& Salaris, M.\ 1998, \memsai, 69, 127

\bibitem[Ciardullo et al.(1990)]{1990BAAS...22.1312C} Ciardullo, R.,
  Jacoby, G.~H., \& Engle, K.~A.\ 1990, \baas 22, 1312

\bibitem[Ciardullo et al.(1989)]{1989ApJ...339...53C} Ciardullo, R.,
  Jacoby, G.~H., Ford, H.~C., et al.\ 1989, \apj, 339, 53

\bibitem[Ciardullo et al.(1993)]{1993ApJ...419..479C} Ciardullo, R.,
  Jacoby, G.~H., \& Tonry, J.~L.\ 1993, \apj, 419, 479

\bibitem[Courtois \& Tully(2012)]{2012ApJ...749..174C} Courtois,
  H.~M., \& Tully, R.~B.\ 2012, \apj, 749, 174

\bibitem[Crandall \& Ratra(2015)]{2015ApJ...815...87C} Crandall, S.,
  \& Ratra, B.\ 2015, \apj, 815, 87

\bibitem[de Grijs \& Bono(2014)]{2014AJ....148...17D} de Grijs, R., \&
  Bono, G.\ 2014, \aj, 148, 17 (Paper II)

\bibitem[de Grijs \& Bono(2015)]{2015AJ....149..179D} de Grijs, R., \&
  Bono, G.\ 2015, \aj, 149, 179 (Paper III)

\bibitem[de Grijs \& Bono(2016)]{2016ApJS..227....5D} de Grijs, R., \&
  Bono, G.\ 2016, \apjs, 227, 5 (Paper IV)

\bibitem[de Grijs \& Bono(2017)]{2017ApJS..232...22D} de Grijs, R., \&
  Bono, G.\ 2017, \apjs, 232, 22 (Paper V)

\bibitem[de Grijs \& Bono(2020)]{2020ApJS..246....3D} de Grijs, R., \&
  Bono, G.\ 2020, \apjs, 246, 3 (Paper VI)

\bibitem[de Grijs et al.(2014)]{2014AJ....147..122D} de Grijs, R.,
  Wicker, J.~E., \& Bono, G.\ 2014, \aj, 147, 122 (Paper I)

\bibitem[de Vaucouleurs(1993)]{1993ApJ...415...10D} de Vaucouleurs,
  G.\ 1993, \apj, 415, 10

\bibitem[Dirsch et al.(2003)]{2003AJ....125.1908D} Dirsch, B.,
  Richtler, T., Geisler, D., et al.\ 2003, \aj, 125, 1908

\bibitem[Drinkwater et al.(2001)]{2001ApJ...548L.139D} Drinkwater,
  M.~J., Gregg, M.~D., \& Colless, M.\ 2001, \apjl, 548, L139

\bibitem[Dunn \& Jerjen(2006)]{2006AJ....132.1384D} Dunn, L.~P., \&
  Jerjen, H.\ 2006, \aj, 132, 1384

\bibitem[Feldmeier et al.(2007)]{2007ApJ...657...76F} Feldmeier,
  J.~J., Jacoby, G.~H., \& Phillips, M.~M.\ 2007, \apj, 657, 76

\bibitem[Ferrarese et al.(2000)]{2000ApJ...529..745F} Ferrarese, L.,
  Mould, J.~R., Kennicutt, R.~C., et al.\ 2000, \apj, 529, 745

\bibitem[Forte et al.(2005)]{2005MNRAS.357...56F} Forte, J.~C.,
  Faifer, F., \& Geisler, D.\ 2005, \mnras, 357, 56

\bibitem[Freedman et al.(2001)]{2001ApJ...553...47F} Freedman, W.~L.,
  Madore, B.~F., Gibson, B.~K., et al.\ 2001, \apj, 553, 47

\bibitem[Fukugita et al.(1991)]{1991ApJ...376....8F} Fukugita, M.,
  Okamura, S., Tarusawa, K., et al.\ 1991, \apj, 376, 8

\bibitem[Geisler \& Forte(1990)]{1990ApJ...350L...5G} Geisler, D., \&
  Forte, J.~C.\ 1990, \apjl, 350, L5

\bibitem[G{\'o}mez et al.(2001)]{2001AnA...371..875G} G{\'o}mez, M.,
  Richtler, T., Infante, L., et al.\ 2001, \aap, 371, 875

\bibitem[Gregg(1997)]{1997NewA....1..363G} Gregg, M.~D.\ 1997, \na, 1,
  363

\bibitem[Grillmair et al.(1999)]{1999AJ....117..167G} Grillmair,
  C.~J., Forbes, D.~A., Brodie, J.~P., et al.\ 1999, \aj, 117, 167

\bibitem[Harris(1996)]{1996AJ....112.1487H} Harris, W.~E.\ 1996, \aj,
  112, 1487

\bibitem[Harris et al.(2009)]{2009AJ....137.3314H} Harris, W.~E.,
  Kavelaars, J.~J., Hanes, D.~A., et al.\ 2009, \aj, 137, 3314

\bibitem[Herrnstein et al.(1999)]{1999Natur.400..539H} Herrnstein,
  J.~R., Moran, J.~M., Greenhill, L.~J., et al.\ 1999, \nat, 400, 539

\bibitem[Hilker et al.(1999)]{1999AnAS..138...55H} Hilker, M.,
  Infante, L., \& Richtler, T.\ 1999, \aaps, 138, 55

\bibitem[Hislop et al.(2011)]{2011ApJ...733...75H} Hislop, L., Mould,
  J., Schmidt, B., et al.\ 2011, \apj, 733, 75

\bibitem[Jacoby(1997)]{1997eds..proc..197J} Jacoby, G.~H.\ 1997, in
  The Extragalactic Distance Scale, eds. M. Livio, M. Donahue, \&
  N. Panagia, Cambridge, UK: Cambridge Univ. Press, p. 197

\bibitem[Jang et al.(2018)]{2018ApJ...852...60J} Jang, I.~S., Hatt,
  D., Beaton, R.~L., et al.\ 2018, \apj, 852, 60

\bibitem[Jensen et al.(2003)]{2003ApJ...583..712J} Jensen, J.~B.,
  Tonry, J.~L., Barris, B.~J., et al.\ 2003, \apj, 583, 712

\bibitem[Jensen et al.(1998)]{1998ApJ...505..111J} Jensen, J.~B.,
  Tonry, J.~L., \& Luppino, G.~A.\ 1998, \apj, 505, 111

\bibitem[Jensen et al.(1999)]{1999ApJ...510...71J} Jensen, J.~B.,
  Tonry, J.~L., \& Luppino, G.~A.\ 1999, \apj, 510, 71

\bibitem[Jerjen(2003)]{2003AnA...398...63J} Jerjen, H.\ 2003, \aap,
  398, 63

\bibitem[Jord{\'a}n et al.(2006)]{2006ApJ...651L..25J} Jord{\'a}n, A.,
  McLaughlin, D.~E., C{\^o}t{\'e}, P., et al.\ 2006, \apjl, 651, L25

\bibitem[Kashikawa et al.(1998)]{1998ApJ...500..750K} Kashikawa, N.,
  Sekiguchi, M., Doi, M., et al.\ 1998, \apj, 500, 750

\bibitem[Kavelaars et al.(2000)]{2000ApJ...533..125K} Kavelaars,
  J.~J., Harris, W.~E., Hanes, D.~A., et al.\ 2000, \apj, 533, 125

\bibitem[Kennicutt et al.(1998)]{1998ApJ...498..181K} Kennicutt,
  R.~C., Stetson, P.~B., Saha, A., et al.\ 1998, \apj, 498, 181

\bibitem[Kissler-Patig et al.(1997)]{1997AnA...319..470K}
  Kissler-Patig, M., Kohle, S., Hilker, M., et al.\ 1997, \aap, 319,
  470

\bibitem[Kohle et al.(1996)]{1996AnA...309L..39K} Kohle, S.,
  Kissler-Patig, M., Hilker, M., et al.\ 1996, \aap, 309, L39

\bibitem[Lee \& Jang(2016)]{2016ApJ...819...77L} Lee, M.~G., \& Jang,
  I.~S.\ 2016, \apj, 819, 77

\bibitem[Lindblad(1999)]{1999AnARv...9..221L} Lindblad, P.~O.\ 1999,
  \aapr, 9, 221

\bibitem[Liu \& Graham(2001)]{2001ApJ...557L..31L} Liu, M.~C., \&
  Graham, J.~R.\ 2001, \apjl, 557, L31

\bibitem[Liu et al.(2002)]{2002ApJ...564..216L} Liu, M.~C., Graham,
  J.~R., \& Charlot, S.\ 2002, \apj, 564, 216

\bibitem[{\L}okas \& Mamon(2003)]{2003MNRAS.343..401L} {\L}okas,
  E.~L., \& Mamon, G.~A.\ 2003, \mnras, 343, 401

\bibitem[Madore et al.(1996)]{1996AAS...18910804M} Madore, B.~F.,
  Freedman, W.~L., Kennicutt, R.~C., et al.\ 1996, Am.
  Astron. Soc. Mtg Abstr. 189, 108.04

\bibitem[Madore et al.(1998)]{1998Natur.395...47M} Madore, B.~F.,
  Freedman, W.~L., Silbermann, N., et al.\ 1998, \nat, 395, 47

\bibitem[Madore et al.(1999)]{1999ApJ...515...29M} Madore, B.~F.,
  Freedman, W.~L., Silbermann, N., et al.\ 1999, \apj, 515, 29

\bibitem[Masters et al.(2006)]{2006ApJ...653..861M} Masters, K.~L.,
  Springob, C.~M., Haynes, M.~P., et al.\ 2006, \apj, 653, 861

\bibitem[McMillan et al.(1993)]{1993ApJ...416...62M} McMillan, R.,
  Ciardullo, R., \& Jacoby, G.~H.\ 1993, \apj, 416, 62

\bibitem[Neill et al.(2014)]{2014ApJ...792..129N} Neill, J.~D.,
  Seibert, M., Tully, R.~B., et al.\ 2014, \apj, 792, 129

\bibitem[Ostrov et al.(1998)]{1998AJ....116.2854O} Ostrov, P.~G.,
  Forte, J.~C., \& Geisler, D.\ 1998, \aj, 116, 2854

\bibitem[Prosser et al.(1999)]{1999ApJ...525...80P} Prosser, C.~F.,
  Kennicutt, R.~C., Bresolin, F., et al.\ 1999, \apj, 525, 80

\bibitem[Richtler et al.(2000)]{2000fepc.conf..259R} Richtler, T.,
  Drenkhahn, G., G{\'o}mez, M., et al.\ 2000, in From Extrasolar
  Planets to Cosmology: The VLT Opening Symposium, eds J. Bergeron \&
  A. Renzini, p. 259

\bibitem[Riess et al.(2019)]{2019ApJ...876...85R} Riess, A.~G.,
  Casertano, S., Yuan, W., et al.\ 2019, \apj, 876, 85

\bibitem[Riess et al.(2009)]{2009ApJ...699..539R} Riess, A.~G., Macri,
  L., Casertano, S., et al.\ 2009, \apj, 699, 539

\bibitem[Riess et al.(2016)]{2016ApJ...826...56R} Riess, A.~G., Macri,
  L.~M., Hoffmann, S.~L., et al.\ 2016, \apj, 826, 56

\bibitem[Rood \& Williams(1993)]{1993MNRAS.263..211R} Rood, H.~J., \&
  Williams, B.~A.\ 1993, \mnras, 263, 211

\bibitem[Russell(2004)]{2004ApJ...607..241R} Russell, D.~G.\ 2004,
  \apj, 607, 241

\bibitem[Russell(2009)]{2009JApA...30...93R} Russell, D.~G.\ 2009,
  J. Astrophys. Astron., 30, 93

\bibitem[Salaris \& Cassisi(1998)]{1998MNRAS.298..166S} Salaris, M.,
  \& Cassisi, S.\ 1998, \mnras, 298, 166

\bibitem[Schaefer(2008)]{2008AJ....135..112S} Schaefer, B.~E.\ 2008,
  \aj, 135, 112

\bibitem[Shanks(1997)]{1997MNRAS.290L..77S} Shanks, T.\ 1997, \mnras,
  290, L77

\bibitem[Silbermann et al.(1999)]{1999ApJ...515....1S} Silbermann,
  N.~A., Harding, P., Ferrarese, L., et al.\ 1999, \apj, 515, 1

\bibitem[Sorce et al.(2013)]{2013ApJ...765...94S} Sorce, J.~G.,
  Courtois, H.~M., Tully, R.~B., et al.\ 2013, \apj, 765, 94

\bibitem[Sorce et al.(2012)]{2012ApJ...758L..12S} Sorce, J.~G., Tully,
  R.~B., \& Courtois, H.~M.\ 2012, \apjl, 758, L12

\bibitem[Sorce et al.(2014)]{2014MNRAS.444..527S} Sorce, J.~G., Tully,
  R.~B., Courtois, H.~M., et al.\ 2014, \mnras, 444, 527

\bibitem[Springob et al.(2007)]{2007ApJS..172..599S} Springob, C.~M.,
  Masters, K.~L., Haynes, M.~P., et al.\ 2007, \apjs, 172, 599

\bibitem[Tammann \& Sandage(1999)]{1999ASPC..167..204T} Tammann,
  G.~A., \& Sandage, A.\ 1999, in Harmonizing Cosmic Distance Scales
  in a Post-hipparcos Era, eds D. Egret \& A. Heck, ASP
  Conf. Ser. 167, 204

\bibitem[Terlevich et al.(1999)]{1999MNRAS.310..445T} Terlevich,
  A.~I., Kuntschner, H., Bower, R.~G., et al.\ 1999, \mnras, 310, 445

\bibitem[Thomsen et al.(1997)]{1997ApJ...483L..37T} Thomsen, B., Baum,
  W.~A., Hammergren, M., et al.\ 1997, \apjl, 483, L37

\bibitem[Tikhonov \& Galazutdinova(2011)]{2011AstL...37..766T}
  Tikhonov, N.~A., \& Galazutdinova, O.~A.\ 2011, Astron. Lett., 37,
  766

\bibitem[Tonry(1991a)]{1991BAAS...23..956T} Tonry, J.~L.\ 1991a,
  \baas, 23, 956

\bibitem[Tonry(1991b)]{1991ApJ...373L...1T} Tonry, J.~L.\ 1991b,
  \apjl, 373, L1

\bibitem[Tonry(1997)]{1997eds..proc..297T} Tonry, J.~L.\ 1997, in The
  Extragalactic Distance Scale, eds. M. Livio, M. Donahue, \&
  N. Panagia, Cambridge, UK: Cambridge Univ. Press, p.  297

\bibitem[Tonry et al.(1997)]{1997ApJ...475..399T} Tonry, J.~L.,
  Blakeslee, J.~P., Ajhar, E.~A., et al.\ 1997, \apj, 475, 399

\bibitem[Tonry et al.(2000)]{2000ApJ...530..625T} Tonry, J.~L.,
  Blakeslee, J.~P., Ajhar, E.~A., et al.\ 2000, \apj, 530, 625

\bibitem[Tonry et al.(2001)]{2001ApJ...546..681T} Tonry, J.~L.,
  Dressler, A., Blakeslee, J.~P., et al.\ 2001, \apj, 546, 681

\bibitem[Tully(1998)]{1998MmSAI..69..237T} Tully, R.~B.\ 1998,
  \memsai, 69, 237

\bibitem[Tully \& Courtois(2012)]{2012ApJ...749...78T} Tully, R.~B.,
  \& Courtois, H.~M.\ 2012, \apj, 749, 78

\bibitem[Tully \& Pierce(2000)]{2000ApJ...533..744T} Tully, R.~B., \&
  Pierce, M.~J.\ 2000, \apj, 533, 744

\bibitem[Tully et al.(2013)]{2013AJ....146...86T} Tully, R.~B.,
  Courtois, H.~M., Dolphin, A.~E., et al.\ 2013, \aj, 146, 86

\bibitem[Tully et al.(2009)]{2009AJ....138..323T} Tully, R.~B., Rizzi,
  L., Shaya, E.~J., et al.\ 2009, \aj, 138, 323

\bibitem[van den Bergh(1994)]{1994PASP..106.1113V} van den Bergh,
  S.\ 1994, \pasp, 106, 1113

\bibitem[van Dokkum et al.(2015)]{2015ApJ...798L..45V} van Dokkum,
  P.~G., Abraham, R., Merritt, A., et al.\ 2015, \apjl, 798, L45

\bibitem[Villegas et al.(2010)]{2010ApJ...717..603V} Villegas, D.,
  Jord{\'a}n, A., Peng, E.~W., et al.\ 2010, \apj, 717, 603

\bibitem[Watanabe et al.(2001)]{2001ApJ...555..215W} Watanabe, M.,
  Yasuda, N., Itoh, N., et al.\ 2001, \apj, 555, 215

\bibitem[Whitmore(1997)]{1997eds..proc..254W} Whitmore, B.~C.\ 1997,
  in The Extragalactic Distance Scale, eds. M. Livio, M. Donahue, \&
  N. Panagia, Cambridge, UK: Cambridge Univ. Press, p. 254

\end{thebibliography}
\end{document}